\documentclass[lettersize,journal]{IEEEtran}
\usepackage{amsmath,amsfonts}
\usepackage{algorithmic}
\usepackage{array}
\usepackage[caption=false,font=normalsize,labelfont=sf,textfont=sf]{subfig}
\usepackage{textcomp}
\usepackage{stfloats}
\usepackage{url}
\usepackage{verbatim}
\usepackage{graphicx}
\def\BibTeX{{\rm B\kern-.05em{\sc i\kern-.025em b}\kern-.08em
    T\kern-.1667em\lower.7ex\hbox{E}\kern-.125emX}}
\usepackage{balance}
\usepackage{cite}
\usepackage{amssymb}
\usepackage{amsmath}
\usepackage{multirow}
\usepackage{booktabs} 

\begin{document}
\title{STEPC: A Multi-energy Nonuniform Response Calibration Framework for Photon-Counting Micro-CT in Multi-material Imaging}
\author{Enze Zhou, Wenjian Li, Wenting Xu, Yuwei Lu, Shangbin Chen, Shaoyang Wang, Gang Zheng, Tianwu Xie, and Qian Liu
\thanks{This work was supported by National Natural Science Foundation of China (32227801). (Corresponding author: Qian Liu; Tianwu Xie.)}
\thanks{E.Zhou, W.Li, W.Xu, Y.Lu and S.Chen are with the MOE Key Laboratory for Biomedical Photonics, Wuhan National Laboratory for Optoelectronics, Huazhong University of Science and Technology, Wuhan 430074, China (e-mail: enzezhou@hust.edu.cn; liwenjian@hust.edu.cn; xuwenting@hust.edu.cn; luyw@hust.edu.cn; sbchen@hust.edu.cn;).}
\thanks{Q.Liu and S.Wang are with the State Key Laboratory of Digital Medical Engineering, School of Biomedical Engineering, Hainan University, Sanya, 572000, China (e-mail: qliu@hainanu.edu.cn; wangshaoyang@hainanu.edu.cn).}
\thanks{T.Xie is with the Institute of Radiation Medicine, Fudan University, Shanghai, China (e-mail: tianwuxie@fudan.edu.cn).}
\thanks{G.Zheng is with the School of Electronic Information and Communications, Huazhong University of Science and Technology, Wuhan, 430074, China (e-mail: ZhengGang@hust.edu.cn).}
}
\markboth{IEEE TRANSACTIONS ON INSTRUMENTATION AND MEASUREMENT}%
{STEPC: A Multi-energy Nonuniform Response Calibration Framework for Photon-Counting Micro-CT in Multi-material Imaging}

\maketitle

\begin{abstract}

Photon-counting computed tomography has demonstrated significant advancements in recent years; however, micro photon-counting CT (Micro-PCCT) systems are still limited by pixel-wise detector response nonuniformity, which degrades measurement uniformity across detector pixels and commonly produces ring artifacts in reconstructed images. Existing calibration methods exhibit limited generalizability in complex multi-material scenarios, such as contrast-enhanced imaging. This study introduces a Signal-to-Nonuniformity Error Polynomial Calibration (STEPC) framework based on measurement nonuniformity error modeling to address this issue. STEPC first fits multi-energy projections using a 2D polynomial surface to generate ideal references, then applies a nonlinear multi-energy polynomial model to predict and correct pixel-wise nonuniformity errors. The model is calibrated using homogeneous slab phantoms of different materials, including PMMA, aluminum, and iodinated contrast agents, enabling correction for both non-contrast and contrast-enhanced imaging. Experiments were performed on a custom Micro-PCCT system with phantoms and mouse. Correction performance of STEPC was evaluated using the mean local standard deviation (MLSD) in the projection domain and the ring artifact deviation (RAD) on the reconstructed images. Compared with existing methods, STEPC achieved an average MLSD reduction of at least 21.58\% and reduced RAD by at least 14.18\%, consistently yielding the best performance in both non-contrast and contrast-enhanced scenarios. Furthermore, STEPC can be readily extended to compensate for beam hardening effects within the same calibration framework. Quantitative material decomposition results indicate that the proposed method preserves measurement accuracy across different basis materials. These results demonstrate that STEPC provides a robust and practical calibration framework that significantly improves detector measurement uniformity in Micro-PCCT systems for multi-material spectral imaging.

\end{abstract}


\section{Introduction}
\label{sec:introduction}

\IEEEPARstart{P}{hoton}-counting computed tomography (PCCT) has gained considerable attention due to its ability to directly measure photon counts in multiple energy windows, enabling ultra-high spatial resolution imaging, material decomposition, reduced beam hardening, and K-edge imaging\cite{van2023photon,schwartz2025photon,schwartz2023exploiting,fan2025beam,badea2019functional}. Despite the numerous advantages of PCCT, its image quality critically depends on the uniformity of detector pixel responses. In practical systems, variations in detector crystals, electronic gain, and readout circuitry result in energy-dependent pixel response nonuniformity and threshold bias. These effects are particularly pronounced in Micro-PCCT systems, where reduced pixel sizes exacerbate pixel-to-pixel response variability. Such detector measurement nonuniformities propagate through the image reconstruction process and manifest as ring artifacts in the reconstructed images~\cite{zhou2025cone,taguchi2020multienergy,feng2021experimental,chen2025energy,yang2024development}.

Various strategies have been proposed to suppress or correct pixel-wise detector nonuniformity in photon-counting CT, which can be broadly categorized into three main approaches: (1) hardware calibration or control methods, (2) image processing-based methods, and (3) phantom-based measurement calibration. These approaches are often complementary and can be combined to achieve optimal correction performance. Hardware calibration focuses on the detector's energy threshold calibration, which directly calibrates detector's energy thresholds using monochromatic synchrotron sources, radioactive isotopes, X-ray fluorescence, or K-edge materials~\cite{vespucci2018robust,rodesch2023comparison}. However, this approach typically requires monochromatic sources and detector repositioning, which is not readily available for most researchers. Hardware control strategies mitigate nonuniformity by randomly moving the sample~\cite{hubert2018efficient} or the detector~\cite{zhu2013micro}, though this increases scan time and demands precise mechanical control. Helical scanning, widely used in clinical CT, can also convert ring-like artifacts to helix-like artifacts, thereby substantially reducing visible rings~\cite{pelt2018ring}.

Image processing-based methods include projection-domain stripe suppression~\cite{munch2009stripe}, pixel-wise gain correction~\cite{ketcham2006new,undefined2020ring,guo2022nonlinear}, and ring removal in the reconstructed image by transforming it into the polar domain, performing stripe suppression, and then converting back to cartesian coordinates~\cite{yan2016variation,wu2019removing,yang2020post}. Dual-domain approaches have also been developed to enhance correction capability~\cite{sun2025ring}. These methods are easy to implement and have been widely used; however, their correction capability is often limited, especially for low-frequency concentric artifacts~\cite{feng2021experimental} or severe nonlinear detector responses~\cite{vaagberg2017removal}, and they may even introduce new artifacts in complex object structures~\cite{pelt2018ring}. Recently, deep learning has shown increasing promise for ring artifact suppression. Fang \textit{et al.}~\cite{fang2020removing} employed deep learning methods to remove ring artifacts by comparing models trained in the image domain, projection domain, and polar coordinate system, and further proposed a comprehensive model that effectively integrates multi-domain information for improved ring artifact suppression. Lv \textit{et al.}~\cite{lv2020image} applied deep learning for simultaneous denoising and ring artifact correction in PCCT, while Yuan \textit{et al.}~\cite{yuan2021deep} employed CNN and RNN architectures for ring artifact removal. Hein \textit{et al.}~\cite{hein2023ring} integrated a CNN with a spectral loss to improve correction performance in PCCT, and Sha \textit{et al.}~\cite{sha2024removing} proposed a dual-domain information fusion method using Transformer with a customized polar coordinate loss. Su \textit{et al.}~\cite{su2025ring} designed a neural network framework that employs multilevel discrete wavelet transform (DWT) to achieve independent ring artifacts correction and interaction guidance of global and local features. Despite their effectiveness, these methods rely heavily on large paired training datasets, making practical deployment challenging. To address this, Hein \textit{et al.}~\cite{hein2025syn2real} further introduced a “Syn2Real” pipeline that efficiently synthesizes realistic ring artifacts directly in the image domain.

Phantom-based measurement calibration, which uses calibration scans of known material phantoms to learn a mapping from measured signals to ideal responses~\cite{jakubek2007data,persson2012framework,inkinen2020calibration,lee2025pixel}. Compared with hardware-based methods, it is easier to implement, and unlike image processing approaches, it directly models detector nonuniformity characteristics from the system itself. Consequently, it serves as a key step in many CT reconstruction pipelines for effectively suppressing ring artifacts. Among phantom-based measurement calibration methods, traditional single-energy correction techniques~\cite{seibert1998flat,jakubek2007data,juntunen2019framework,feng2021Overcoming} can improve uniformity in single-material scenarios but perform poorly when multiple materials (e.g., soft tissue, bone, or iodine contrast agents) are present. Material decomposition methods~\cite{wang2008uniformity,alvarez2011estimator,inkinen2020calibration} could address multi-material calibration, but their practicality is restricted by current photon-counting detectors (PCDs) limited energy thresholds (typically $\leq 2$), restricting them to dual-material decomposition in the projection domain. Although multi-material decomposition can be achieved by introducing additional constraints~\cite{lee2014quantitative}, such methods rely heavily on prior information and may introduce new biases. These limitations make existing techniques inadequate for complex tasks such as contrast-enhanced imaging. Furthermore, most of these methods require precise and time-consuming phantom calibration procedures. A recently proposed off-center water phantom calibration technique reduces this complexity~\cite{lee2025calibration}; however, it remains limited in handling multi-material scenarios.

In this study, we propose the Signal-to-Nonuniformity  Error Polynomial Calibration (STEPC), a phantom-based calibration framework that characterizes and corrects material-dependent nonuniform response of PCDs in Micro-CT systems. STEPC models detector nonuniformity by attributing errors to intrinsic detector non-ideal factors and material-dependent X-ray attenuation. A multi-energy polynomial error model incorporates the attenuation characteristics of different materials to predict and correct pixel-wise measurement nonuniformity error. The effectiveness of this method is validated on a custom Micro-PCCT system using phantoms with different materials as well as mouse imaging with and without contrast agents. STEPC remains effective for multi-material imaging scenarios even with only two energy thresholds, and can also be readily extended to compensate for beam hardening. This approach provides a more accurate and comprehensive framework for modeling and calibrating photon-counting detector nonuniformity, and has the potential to serve as a general-purpose calibration framework for Micro-PCCT systems.

The rest of this article is organized as follows. Section~\ref{sec:relatedworks} reviews related work on phantom-based calibration methods and introduces several baseline methods for comparison. Section ~\ref{sec:methods} presents the physical model of detector nonuniformity error and describes the steps of STEPC, along with its extension to beam hardening correction. Section~\ref{sec:experiments} details the experimental setup and evaluation metrics. Section~\ref{sec:results} presents the main results from phantom and mouse imaging. Section~\ref{sec:discussion} discusses the advantages, limitations, and potential future directions of the proposed method. Finally, Section~\ref{sec:conclusion} concludes the article.

\section{Related Works}
\label{sec:relatedworks}

Phantom-based measurement calibration methods typically involve three steps: (1) acquiring all possible incident spectra projections using various homogeneous slab phantoms, (2) calculating the ideal reference projections of the phantom, and (3) fitting a parametric model to map the measured signals to the reference ideal signals, which is then used for correcting actual object imaging. The main differences among these methods lie in the choice of incident spectra, and the form of the ideal signal and correction model.

Typically, different incident spectra are obtained using homogeneous phantoms with different material (such as PMMA and AL) and thickness. The reference signal is commonly selected as either a single pixel value or the mean value~\cite{seibert1998flat,persson2012framework}. Alternative approaches include: (1) 1D polynomial fitting of projections to account for spatial variations in X-ray intensity or phantom thickness~\cite{altunbas2014reduction}. (2) directly using measured thickness values as reference (common in material decomposition methods)~\cite{jakubek2007data,lee2025pixel,juntunen2019framework,feng2021Overcoming}, and (3) simulation of ideal signals using known geometry, source spectra, and detector response functions~\cite{ding2012imagebased,schirra2014towards,liu2015spectral}.

Correction models are generally divided into traditional single-energy and multi-energy approaches. Single-energy methods perform energy-bin-wise corrections using linear~\cite{seibert1998flat}, exponential~\cite{jakubek2007data}, or empirical nonlinear functions~\cite{ding2012imagebased}. While these methods work well for single-component objects, their correction accuracy degrades significantly for multi-material cases involving both soft tissue and bone. In contrast, multi-energy approaches leverage richer spectral information to enable more effective multi-material correction~\cite{persson2012framework,alvarez2011estimator,inkinen2020calibration}. In the following, we introduce two representative single-energy methods: Flat-field Correction (FF)~\cite{moy1999how} and Signal-to-Equivalent Thickness Calibration (STC)~\cite{jakubek2007data}, and two multi-energy methods: Affine Transformation Calibration (ATC)~\cite{persson2012framework} and Polynomial Material Decomposition Calibration (PMDC)~\cite{alvarez2011estimator,inkinen2020calibration}, which will serve as comparative baselines in our study.

\begin{figure}[!t]
    \centering
    \centerline{\includegraphics[width=\columnwidth]{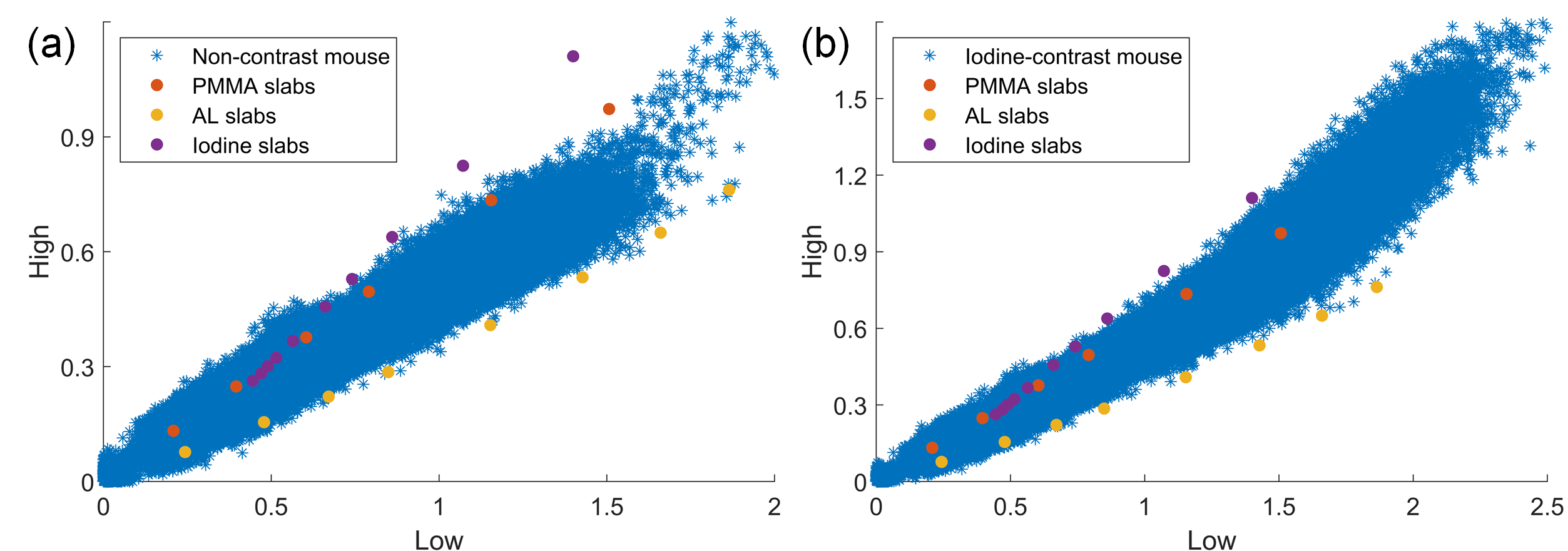}}
    \caption{Distribution of spectral projection values for the non-contrast (a) and iodine-contrast (b) mouse scans.}
    \label{fig:fig1}
\end{figure}

\textbf{(1) Flat-field Correction (FF)}

Flat-field correction is the most common approach in energy-integrating CT systems. It includes dark-field (offset) and gain correction~\cite{moy1999how}. The corrected signal is computed as:
\begin{equation}
    I^{\mathrm{FF}} = \frac{N - B}{N_{\mathrm{air}} - B_{\mathrm{air}}}
\end{equation}
where $B$ and $B_{\mathrm{air}}$ are the dark-field signals for object and air scans, respectively, and $N$ and $N_{\mathrm{air}}$ are the raw counts of object and air scanning. In PCCT, the energy threshold can suppress electronic noise, $B \approx 0$ and $B_{\mathrm{air}} \approx 0$, yielding:
\begin{equation}
    I_E^{\mathrm{FF}} = \frac{N_E}{N_{E,\mathrm{air}}}
\end{equation}
where $E$ indexes the energy bin. The negative logarithmic transform is then applied for reconstruction:
\begin{equation}
    P_E^{\mathrm{FF}} = -\log(I_E^{\mathrm{FF}})
\end{equation}

\textbf{(2) Signal-to-Equivalent Thickness Calibration (STC)}

STC calibrates detector response using varying thicknesses of a single calibration material, typically PMMA for soft-tissue equivalence~\cite{jakubek2007data}. For each energy bin $E$, an exponential model is fit:
\begin{equation}
    N_E = C_E e^{A_E T_E}
\end{equation}
where $N_E$ is the detected photon count, $T_E$ is the material thickness, and $A_E$, $C_E$ are fitted parameters. The corrected equivalent thickness is calculated as:
\begin{equation}
    T_E^{\mathrm{STC}} = \frac{1}{A_E} \ln\left(\frac{N_E}{C_E}\right)
\end{equation}

This corrected thickness can be directly used for image reconstruction or converted into a virtual monoenergetic image for reconstruction.

\textbf{(3) Affine Transformation Calibration (ATC)}

Unlike FF and STC, which calibrate each energy bin data independently, ATC leverages all energy bin counts $N$ jointly to compensate for count loss caused by threshold variations~\cite{persson2012framework}. The corrected counts are given by:
\begin{equation}
    \mathbf{N}^{\mathrm{ATC}} = \mathbf{A}\mathbf{N} + \mathbf{b}
\end{equation}
where $\mathbf{A} \in \mathbb{R}^{K \times K}$ and $\mathbf{b} \in \mathbb{R}^K$, with $K$ denoting the number of energy thresholds. The parameters $\mathbf{A}$ and $\mathbf{b}$ are learned by minimizing the mean squared error of the estimator for reference values $\lambda_i^{\mathrm{ref}}$:
\begin{equation}
    \mathrm{MSE} = \mathbb{E}\left[\left(\sum_{j=1}^{K} A_{ij} \widetilde{N}_j + b_i - \lambda_i^{\mathrm{ref}} \right)^2\right]
\end{equation}
where the expectation value is taken over all possible incident spectra and all possible noise realizations. The reference values $\lambda_i^{\mathrm{ref}}$ typically computed as the average value of projection.

\textbf{(4) Polynomial Material Decomposition Calibration (PMDC)}

Biological objects often consist of multiple materials, such as soft tissue and bone. The PMDC method corrects for multi-material effects by performing projection-domain material decomposition using polynomial fitting~\cite{alvarez2011estimator,inkinen2020calibration}. Given flat-field corrected logarithmic projections $P_E = -\log(I_E^{\mathrm{FF}})$, the thicknesses of two basis materials are estimated as:
\begin{equation}
    T_{M_1}^{\mathrm{PMDC}} = \sum_{i+j \leq p} c_{ij} P_{E_1}^i P_{E_2}^j
\end{equation}
\begin{equation}
    T_{M_2}^{\mathrm{PMDC}} = \sum_{i+j \leq p} d_{ij} P_{E_1}^i P_{E_2}^j
\end{equation}
where $p$ is the polynomial order (\textit{third-order} in this study), and the coefficients \( c_{ij} \) and \( d_{ij} \) are determined from calibration scans using different thicknesses combinations of two basis materials.

Among the above methods, FF and STC calibrate each energy bin independently, ignoring the additional information between energy bins. Moreover, FF does not consider object-induced changes in spectra. STC accounts for spectral changes but assumes a single material, which limits accuracy for biological tissues composed of multiple components (e.g., soft tissue and bone). Some methods extend single-energy correction to multi-material scenarios, such as Feng et al.\cite{feng2021experimental} estimated the thickness of one material, then segmented other components using soft thresholding, but requiring additional steps such as segmentation, forward, and backward projections. Other approaches, PETC\cite{juntunen2019framework} transforms aluminum-equivalent thickness into PMMA-equivalent thickness using bin-wise projection models, but requires iterative per-pixel Gauss-Newton optimization (up to 1000 iterations), resulting in high computational cost. ATC allows for varying incident spectra but assumes a linear detector response, which limits its ability to handle the nonlinear characteristics of practical systems. PMDC improves correction by incorporating multi-material spectral changes and multi-energy bin information, but remains limited by the number of energy thresholds. Most commercial PCCT systems offer only two thresholds, which allows only dual-material decomposition and makes it unsuitable for contrast-enhanced imaging. As shown in Fig.~\ref{fig:fig1}, the incident spectra of non-contrast mouse scans can be modeled using PMMA and aluminum, whereas iodine-enhanced cases require iodine as an additional calibration material. Moreover, PMDC relies on accurate phantom thickness measurements and involves a signal-to-thickness conversion, which introduces additional uncertainty and may amplify noise.

\section{Methods}
\label{sec:methods}
\subsection{Physical basis of nonuniform detector response}

To help explain the principle of the proposed method, the physical origin of the non-uniform detector response in photon-counting detectors (PCDs) is first described in this section. For the $j$th pixel in energy threshold $E_k$, the measured detector counts can be expressed as:

\begin{equation}
N_{k,j} = \int_{E_{k}-\Delta E_{k,j}}^{E_{\max}} \int_0^{E_{\max}} S(E) e^{-\int_{l}{\mu(E, \vec{x}) dl}} R_j(E', E) \, dE \, dE'
\label{eq:eq10}
\end{equation}

where $\Delta E_{k,j}$ denotes the threshold shift of pixel $j$, $S(E)$ is the input X-ray spectrum, $E$ denotes the photon energy, $R_j(E',E)$ represents the pixel-specific photon-counting detector response function, $E'$ denotes the detected energy, $E_{\max}$ is the maximum photon energy determined by the tube voltage. Assuming the existence of an ideal or reference response function $R_j^{\mathrm{ideal}}(E',E)$ that ensures uniformity across pixels, the corresponding ideal detector count is:

\begin{equation}
N_{k,j}^{\mathrm{ideal}} = \int_{E_{k}}^{E_{\max}} \int_0^{E_{\max}} S(E) e^{-\int_{l}{\mu(E, \vec{x}) dl}} R_j^{\mathrm{ideal}}(E', E) \, dE \, dE'
\label{eq:eq11}
\end{equation}

Thus, the nonuniformity error is:

\begin{equation}
\begin{split}
&N_{k,j}^{\mathrm{error}} = N_{k,j} - N_{k,j}^{\mathrm{ideal}} \\
                &\approx \int_{E_{k}}^{E_{\max}} \int_0^{E_{\max}} S(E) e^{-\int_{l}{\mu(E, \vec{x}) dl}} \Delta R_j(E', E) \, dE \, dE' \\
                &+ \Delta E_{k,j} \int_{0}^{E_{\max}} S(E)\, e^{-\int_{l}\mu(E,\vec{x})\,dl}\, R_j^{\mathrm{ideal}}(E_{k},E)\, dE.
\end{split}
\label{eq:eq12}
\end{equation}

where $\Delta R_j(E', E)=R_j(E', E)-R_j^{\mathrm{ideal}}(E', E)$. The detailed derivation of the approximation in Eq.~\eqref{eq:eq12} is provided in Appendix~A. The above equation indicates that $N_{k,j}^{\mathrm{error}}$ arises from two primary sources: the pixel-dependent variation in detector response $\Delta R_j(E',E)$ and the local energy threshold shift $\Delta E_{k,j}$. Both terms are modulated by the energy-dependent incident spectrum $S(E)e^{-\int_{l}\mu(E,\vec{x})\,dl}$, which depends on the material composition and attenuation properties along the X-ray path. Consequently, the nonuniformity error is a nonlinear functional jointly determined by the spectral characteristics of the object and the pixel-wise spectral response of the detector. Therefore, the nonuniformity estimated from single calibration material cannot be directly generalized to other materials or imaging conditions. We will also later prove that it is impossible to estimate the nonuniformity error for multi-material objects using only single-energy information.

\subsection{Signal-to-Nonuniformity  Error Polynomial Calibration (STEPC)}

\begin{figure}[!t]
    \centerline{\includegraphics[width=\columnwidth]{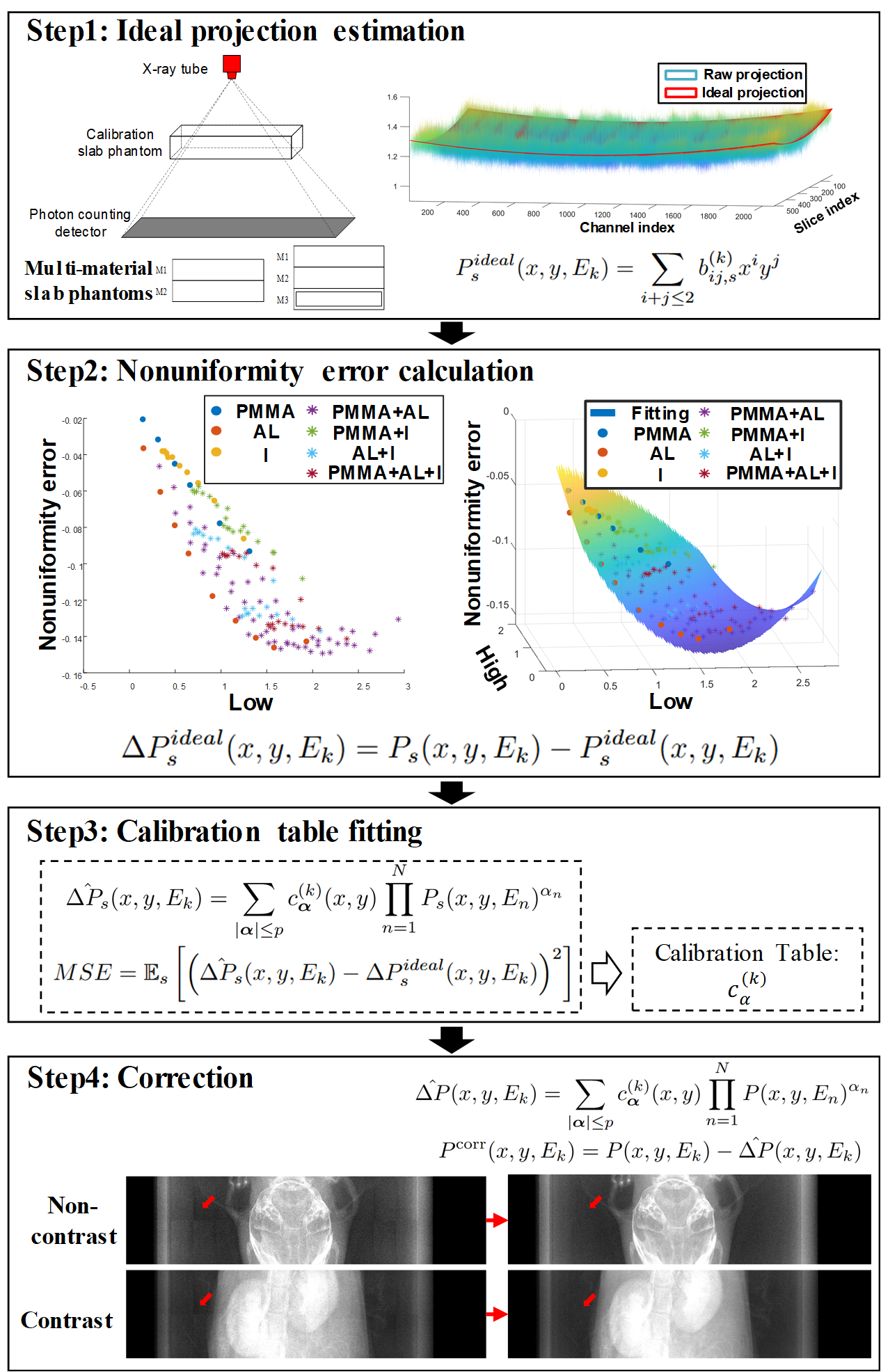}}
    \caption{Workflow of the proposed STEPC method. Step 1: Estimation of ideal projections for calibration slab phantoms; Step 2: Calculation of pixel-wise nonuniformity errors; Step 3: Generation of the calibration table by fitting the empirical multi-energy polynomial model; Step 4: Projection correction for various imaging scenarios (red arrows indicate noticeable pixel response nonuniformity before correction).}
    \label{fig:fig2}
\end{figure}

We propose a STEPC method that directly utilizes multi-energy projection information to predict and correct detector nonuniformity for multi-material imaging. The detailed workflow is illustrated in Fig.~\ref{fig:fig2} and consists of the following steps:

\textbf{Step 1}: Ideal projection estimation. STEPC first measures the transmission through different combinations of multiple material slab phantoms. The log-transformed flat-field corrected projection data are then computed as $P(E_k) = -\log(I_E^{\mathrm{FF}})$ to reduce the influence of X-ray source and detector instabilities. Inspired by Altunbas et al. \cite{altunbas2014reduction}, a second-order polynomial fit is applied to the 2D projection image for each energy bin $E_k$ to obtain the ideal projection:
\begin{equation}
P_s^{\mathrm{ideal}}(x,y,E_k) = \sum_{i+j \leq 2} b_{ij,s}^{(k)} x^i y^j
\end{equation}
Here, $(x,y)$ are the pixel coordinates, and $s$ indexes the calibration combinations of multiple material slabs. The coefficients $b_{ij,s}^{(k)}$ are obtained by minimizing the mean square error between $P_{s}(E_k)$ and $P_s^{\mathrm{ideal}}(E_k)$. We use MATLAB's \texttt{regress} function \cite{mathWorksregress} to solve for these coefficients. 

This low-order polynomial approximation leverages the fact that both the X-ray beam intensity distribution and the transmitted signal through homogeneous or layered slab phantoms vary smoothly across the detector. The spatial dependence introduced by the source heel effect, beam filtration, and oblique incidence geometry changes gradually over the field of view, and thus can be effectively captured by a quadratic surface. As shown in Fig.~\ref{fig:fig2}(Step1), this polynomial fitting compensates for spatial variations in X-ray intensity and angular-dependent transmission thickness in the slab phantom, without requiring precise geometric measurements, thereby simplifying the calibration process.

\textbf{Step 2}: Nonuniformity error calculation. For each pixel and energy channel, the residual between the measured projection and the ideal projection is computed to obtain the pixel-wise nonuniformity error:
\begin{equation}
\Delta P_s^{\mathrm{ideal}}(x,y,E_k) = P_{s}(x,y,E_k) - P_s^{\mathrm{ideal}}(x,y,E_k)
\end{equation}

As shown in Fig.~\ref{fig:fig2}(Step2), the nonuniformity error is dependent on the incident spectrum, which varies with material type and thickness. For single-material cases, single-energy projections can roughly estimate the error. However, for multi-materials, single-energy projections fail to capture the error accurately. In contrast, it illustrates that dual-energy information more effectively characterizes the error surface, allowing accurate modeling of errors caused by multi-materials.

\textbf{Step 3}: Calibration table fitting. As described in Eq.~\ref{eq:eq12}, the nonuniformity error arises from both detector-related non-ideal factors and material-dependent X-ray attenuation. By incorporating projections from all energy channels as input, the model captures spectral variations associated with the attenuation characteristics of different materials. An empirical polynomial model is then employed to implicitly entangle these effects within a unified formulation, enabling more accurate and robust correction in multi-material imaging scenarios. The model is formulated as follows:

\begin{equation}
\hat{\Delta P_{s}}(x,y,E_k) = \sum_{|\boldsymbol{\alpha}| \leq p} c_{\boldsymbol{\alpha}}^{(k)}(x,y) \prod_{n=1}^{N} P_{s}(x,y,E_n)^{\alpha_n}
\end{equation}

where $\boldsymbol{\alpha} = (\alpha_1,\ldots,\alpha_N)$ is the multi-index vector, $|\boldsymbol{\alpha}| = \sum_{n=1}^{N} \alpha_n \leq p$ is the total polynomial degree (\textit{third-order} in this study), $N$ is the number of energy thresholds, and $c_{\boldsymbol{\alpha}}^{(k)}(x,y)$ are the polynomial coefficients for energy channel $E_k$. These coefficients are determined by minimizing the mean square error of the estimator for ideal error $\Delta P_s^{\mathrm{ideal}}$:
\begin{equation}
MSE = \mathbb{E}_{s}\left[ \left(\hat{\Delta P_{s}}(x,y,E_k) - \Delta P_s^{\mathrm{ideal}}(x,y,E_k)\right)^2 \right]
\end{equation}

The expectation is taken over calibration data collected from homogeneous phantoms composed of different materials (e.g., PMMA, Aluminum, Iodixanol) to cover a wide range of possible incident spectra. The coefficients $c_{\boldsymbol{\alpha}}^{(k)}$ can be solved using regression tools such as MATLAB's \texttt{regress} function\cite{mathWorksregress}.

\textbf{Step 4}: Correction. Finally, real object projection data are corrected by directly inputting the raw multi-energy projections into the fitted polynomial model to estimate the nonuniformity error:
\begin{equation}
\hat{\Delta P}(x,y,E_k) = \sum_{|\boldsymbol{\alpha}| \leq p} c_{\boldsymbol{\alpha}}^{(k)}(x,y) \prod_{n=1}^{N} P(x,y,E_n)^{\alpha_n}
\end{equation}
and the corrected projection is obtained as:
\begin{equation}
P^{\mathrm{corr}}(x,y,E_k) = P(x,y,E_k) - \hat{\Delta P}(x,y,E_k)
\end{equation}

Notably, \(c_{\boldsymbol{\alpha}}^{(k)}(x,y)\) indicates that the calibration and correction are independently performed for each detector pixel \((x,y)\) and each energy bin \(E_k\). 

Conventional methods such as STC and PMDC are primarily based on thickness estimation models, whose correction capability in multi-material imaging is inherently limited by the number of effective energy bins. ATC, on the other hand, compensates for count-domain biases induced by energy threshold variations; however, detector nonuniformity in photon-counting CT arises from multiple sources, and a simple affine transformation may be insufficient to capture the resulting nonlinear behavior. In contrast, STEPC directly formulates a nonuniformity error model in the projection domain, attributing projection inconsistencies to the combined effects of detector non-idealities (e.g., threshold variations and response differences) and multi-material X-ray attenuation. By incorporating projections from multiple energy bins, STEPC enriches multi-material attenuation information, while the polynomial model implicitly and effectively entangles detector non-idealities with material-dependent attenuation effects.

Furthermore, STEPC avoids the need for accurate thickness or density measurements of calibration phantoms by using 2D polynomial fitting to generate ideal reference projections. In addition, despite having only two energy thresholds, STEPC can incorporate iodine-based calibration to capture spectral variations from contrast agents, soft tissue, and bone, enabling effective correction in complex imaging scenarios.

\subsection{Extension to Beam Hardening Correction}
\label{sec:BHcorr}

The proposed STEPC framework can be further extended to correct for beam hardening (BH) effects. The overall procedure follows a similar structure to STEPC. In \textbf{Step~1}, the ideal projection $P_s^{\mathrm{ideal}}(x,y,E_k)$ for each calibration slab combination $s$ is obtained as described previously. The difference lies in \textbf{Step~2}, where the beam hardening induced projection error is computed as:
\begin{equation}
\begin{split}
\Delta P_s^{\mathrm{BHideal}}&(x,y,E_k) = \\ 
&P_s^{\mathrm{BHideal}}(x,y,E_k) - P_s^{\mathrm{ideal}}(x,y,E_k)
\end{split}
\end{equation}
where the beam hardening affected ideal projection $P_s^{\mathrm{BHideal}}$ is modeled as:
\begin{equation}
P_s^{\mathrm{BHideal}}(x,y,E_k) = \sum_{m=1}^{M} T_m^{(s)}(x,y)\mu_m(E_k)
\end{equation}
Here, $T_m^{(s)}$ denotes the thickness of the $m$th material in the $s$th slab combination, $M$ is the number of materials, and $\mu_m(E_k)$ represents the effective linear attenuation coefficient at energy bin $E_k$.  
In biomedical imaging, beam hardening is mainly caused by low-density soft tissue and high-density bone, thus PMMA and aluminum ($M=2$) are typically employed for BH calibration. The $\mu_m(E_k)$ values can be estimated via linear fitting from PMMA and aluminum slab measurements.

Using the multi-energy projection information, the beam-hardening correction term is then estimated with a polynomial regression model analogous to STEPC:
\begin{equation}
\begin{split}
\hat{\Delta P}_s^{\mathrm{BHcorr}}&(x,y,E_k) = \\
&\sum_{|\boldsymbol{\alpha}| \leq p} c_{\mathrm{BHcorr},\boldsymbol{\alpha}}^{(k)}(x,y) 
\prod_{n=1}^{N} P_s^{\mathrm{ideal}}(x,y,E_n)^{\alpha_n}
\end{split}
\end{equation}

Here, the polynomial degree $p$ is also set to three, and the coefficients $c_{\mathrm{BHcorr},\boldsymbol{\alpha}}^{(k)}(x,y)$ are determined by minimizing the mean square error over all calibration combinations $s$:
\begin{equation}
\begin{split}
&MSE = \\
&\mathbb{E}_{s}\left[ \left(\hat{\Delta P}_s^{\mathrm{BHcorr}}(x,y,E_k) - 
\Delta P_s^{\mathrm{BHideal}}(x,y,E_k)\right)^2 \right]    
\end{split}
\end{equation}

Finally, beam hardening correction is applied to the nonuniformity corrected projection $P^{\mathrm{corr}}(x,y,E_k)$:
\begin{equation}
\begin{split}
\hat{\Delta P}^{\mathrm{BHcorr}}&(x,y,E_k) = \\
&\sum_{|\boldsymbol{\alpha}| \leq p} c_{\mathrm{BHcorr},\boldsymbol{\alpha}}^{(k)}(x,y) 
\prod_{n=1}^{N} P^{\mathrm{corr}}(x,y,E_n)^{\alpha_n}    
\end{split}
\end{equation}
and the beam hardening corrected projection is obtained as:
\begin{equation}
P^{\mathrm{BHcorr}}(x,y,E_k) = P^{\mathrm{corr}}(x,y,E_k) - \hat{\Delta P}^{\mathrm{BHcorr}}(x,y,E_k)
\end{equation}

It should be noted that the beam hardening correction extension is not inherently coupled with the detector nonuniformity model. It can be applied independently as a standalone correction module and still provides effective beam hardening mitigation. Its applicability to other calibration methods is further demonstrated in Section~\ref{sec:MDQuantitative}.

\begin{figure}[!t]
    \centerline{\includegraphics[width=\columnwidth]{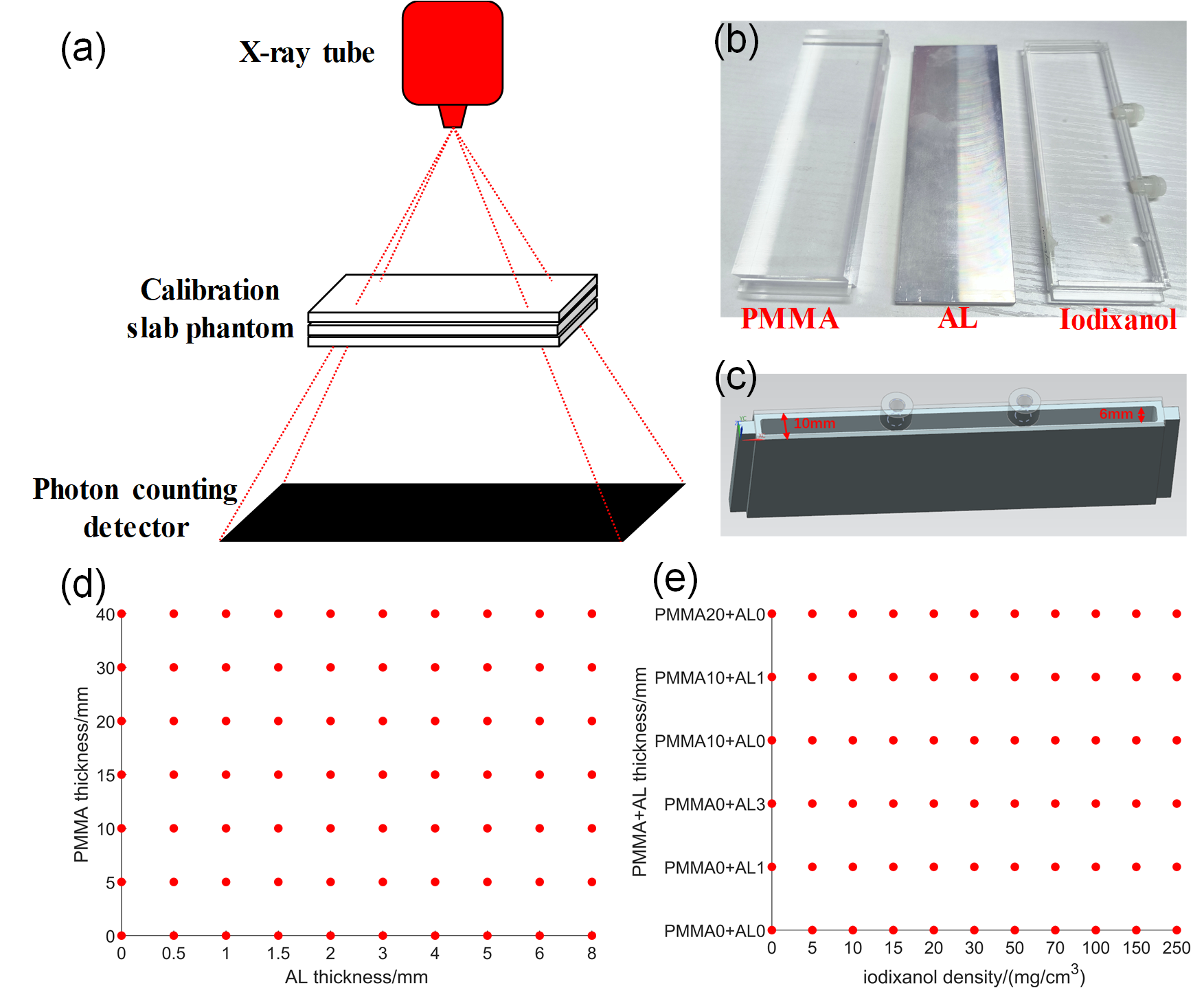}}
    \caption{Schematic of the system geometry and multi-material slab phantoms. (a) Geometry configuration of the calibration setup; (b) three types of slab phantoms: PMMA, aluminum, and iodixanol; (c) schematic of iodixanol phantom container; (d) combinations of PMMA and aluminum slabs with varying thicknesses; (e) combinations of iodixanol solution slabs with PMMA and aluminum slabs.}
    \label{fig:fig3}
\end{figure}

\section{Experiments}
\label{sec:experiments}
\subsection{Micro Photon Counting CT and Calibration Materials}
\label{sec:CalibrationSetting}

The Micro-PCCT system used in our experiments was jointly developed by Hainan University and United Imaging Life Science Instrument (LSI, Wuhan, China). It adopts a translate-rotate architecture where the object remains stationary and the gantry rotates, reducing motion artifacts during imaging. The X-ray source was operated at 80 kV and 200 µA, and a 0.5 mm aluminum filter was used to suppress low-energy photons. The photon counting detector has a resolution of $2063 \times 505$ pixels (after cropping peripheral invalid pixels), each with a 100~$\mu$m~$\times$~100~$\mu$m pixel size, and supports two independently adjustable energy thresholds. The detector thresholds were set to 15 keV and 30 keV. Each energy channel uses a 12-bit counter capable of recording up to 4096 photons per acquisition. The exposure time per projection was 70 ms to keep photon counts within the detector's dynamic range. Each acquisition thus provides three energy bins: \textbf{Total} (15-80 keV) and \textbf{High} (30-80 keV), with the \textbf{Low} (15-30 keV) bin derived by subtraction: $\mathrm{Low} = \mathrm{Total} - \mathrm{High}$. 

\begin{figure}[!t]
    \centerline{\includegraphics[width=\columnwidth]{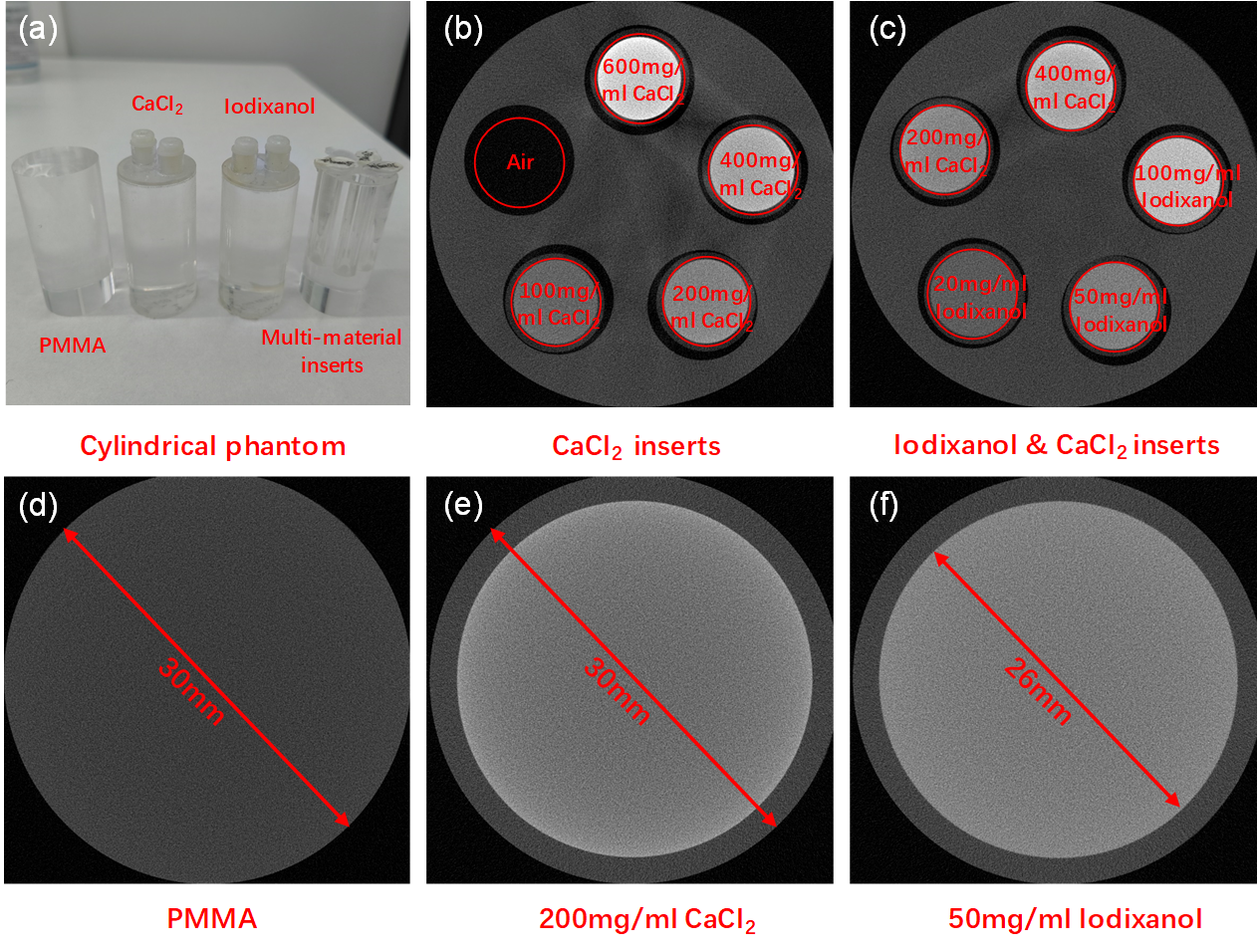}}
    \caption{multi-material cylindrical phantoms. (a) Actual pictures of cylindrical phantoms; (b) CaCl$_2$ inserts with concentrations of 100, 200, 400, and 600 mg/mL insert in a PMMA holder; (c) Iodixanol and CaCl$_2$ inserts (iodixanol: 20, 50, 100 mg/mL; CaCl$_2$: 200, 400 mg/mL); (d) PMMA only; (e) 200 mg/mL CaCl$_2$ (2mm thickness PMMA cylindrical holder); (f) 50 mg/mL iodixanol.}
    \label{fig:fig4}
\end{figure}

Three types of slab phantoms were designed, as shown in Fig.~\ref{fig:fig3}(b), including PMMA slabs with thicknesses of 0, 5, 10, 15, 20, 30, and 40 mm; aluminum slabs with thicknesses of 0, 0.5, 1, 1.5, 2, 3, 4, 5, 6, and 8 mm; and iodixanol solution slabs with concentrations of 0, 5, 10, 15, 20, 30, 50, 70, 100, 150, and 250 mg/cm\textsuperscript{3}, sealed in 2 mm thick PMMA containers with a 6 mm solution core (Fig.~\ref{fig:fig3}(c)). The slab phantoms were positioned approximately parallel to the detector plane, as illustrated in Fig.~\ref{fig:fig3}(a). Two imaging scenarios were considered. For the non-contrast scenario, calibration was performed using combinations of PMMA and aluminum slabs, as shown in Fig.~\ref{fig:fig3}(d). For the iodine-enhanced scenario, the same PMMA and aluminum slab combinations were used, and additional calibration was performed by incorporating iodixanol solutions combined with PMMA and aluminum slabs ({0,0}, {0,1}, {0,3}, {10,0}, {10,1}, and {20,0}), as illustrated in Fig.~\ref{fig:fig3}(e). For each slab combination, 600 projection frames were acquired and averaged to reduce noise.

\subsection{Phantom and Mouse Imaging}

\paragraph{Cylindrical Phantom Scanning}
To evaluate correction performance under different conditions, five cylindrical phantoms were scanned (Fig.~\ref{fig:fig4}): PMMA-only cylinders, 200 mg/mL CaCl$_2$ cylinders (PMMA container with 2 mm thickness), 50 mg/mL Iodixanol cylinder (PMMA container with 2 mm thickness), cylinders with inserts of CaCl$_2$ at $\{100, 200, 400, 600\}$ mg/mL, cylinders with inserts of CaCl$_2$ $\{200, 400\}$ mg/mL and iodixanol $\{20, 50, 100\}$ mg/mL. All phantoms had an outer diameter of 30 mm, and the inserts diameter is 8 mm. Projections were acquired in “Continuous” mode with 1440 views per rotation (0.25° per view), a field of view (FOV) of 50 mm, a source-to-isocenter distance (SID) of 90 mm, and a source-to-detector distance (SDD) of 325 mm. All other scan settings were the same as in the calibration step.

\paragraph{Mouse Scanning}
To validate correction performance in vivo, two C57BL mice (10 weeks old, $\sim$23 g) were scanned. One mouse underwent a non-contrast head scan, while the other received an intravenous injection of 0.3 mL iodixanol (300 mg/ml) via the tail vein and was scanned in the abdominal kidney region. Both mice were euthanized through intraperitoneal anesthesia prior to scanning to eliminate motion artifacts. Imaging parameters were the same as those used for the cylindrical phantom, except for FOV = 35 mm, SID = 74 mm, SDD = 325 mm.

All animal experiments were conducted using mice maintained in SPF animal facilities. Animal care and experimental protocols were approved by the Institutional Animal Care and Use Committee (IACUC) and the Ethical Committee of Animal Experiments of the School of Biomedical Engineering at Hainan University (Approval number: HNUAUCC-2022-00091).
\subsection{Data Processing and Reconstruction}
Detector nonuniformity correction was performed in the projection domain. The detector comprises $16 \times 2$ tiles, separated by gap pixels and affected by defective pixels. These gap and defective pixels were pre-identified and corrected via linear interpolation. Image reconstruction was carried out using the open-source TIGRE toolbox~\cite{biguri2016TIGRE}, employing the FDK algorithm with Hann window filtering. The reconstructed image matrix size was $1529\times1529\times400$. The reconstruction FOV was $50~\mathrm{mm}\times50~\mathrm{mm}\times10~\mathrm{mm}$ for the cylindrical phantoms and $35~\mathrm{mm}\times35~\mathrm{mm} \times8~\mathrm{mm}$ for the mice. Images were converted to Hounsfield Units (HU) using a water phantom for calibration. All processing was performed in MATLAB R2024a.
\begin{figure}[!t]
    \centering
    \includegraphics[width=8cm]{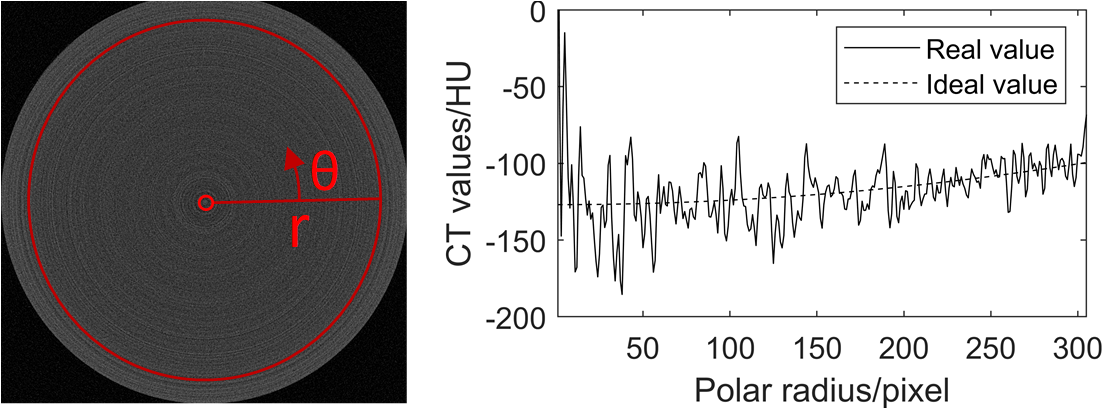}
    \caption{Illustration of the Ring Artifact Deviation (RAD) calculation. \textbf{Left}: Reconstructed image of a PMMA phantom shown in polar coordinates \((r, \theta)\). \textbf{Right}: Angularly averaged intensity profile. A second-order polynomial fit is used to generate the ideal baseline, which is subtracted from the real averaged profile to remove cupping artifacts.}
    \label{fig:fig5}
\end{figure}
\subsection{Baselines}
We compared our STEPC method with four baselines: FF, STC, ATC, and PMDC. Since these methods work in different domains, appropriate post-processing was applied to ensure a fair comparison:
\begin{itemize}
    \item \textbf{FF:} Logarithmic projections were directly computed as $P_E^{\mathrm{FF}} = -\log(I_E^{\mathrm{FF}})$.
    \item \textbf{STC:} Projections were computed using the calibrated thickness model: $P_E^{\mathrm{STC}} = -\log\left(\frac{N_E^{\mathrm{STC}}}{\bar{N}_{E,\mathrm{air}}}\right)$, where $N_E^{\mathrm{STC}} = \bar{C}_E e^{\bar{A}_E T_E}$ and $\bar{C}_E$, $\bar{A}_E$ are average calibration coefficients obtained across all pixels.
    \item \textbf{ATC:} Instead of using mean counts as described in Ref.~\cite{persson2012framework}, the same second-order 2D polynomial fit was applied to generate reference photon counts. After ATC correction in the counting domain, logarithmic air normalization was performed: $P_E^{\mathrm{ATC}} = -\log\left(\frac{N_E^{\mathrm{ATC}}}{\acute{N}_{E,\mathrm{air}}}\right)$. $\acute{N}_{E,\mathrm{air}}$ was also obtained by applying a second-order 2D polynomial fit to the air scan projection.
    \item \textbf{PMDC:} For each energy bin $E(k)$, the following polynomial forward model was fitted:
    \begin{equation}
    P_{E(k)} = \sum_{1 \le i+j \le p} c_{ij}^{(k)} (T_{\mathrm{PMMA}})^i (T_{\mathrm{Al}})^j,
    \label{eq:eq25}
    \end{equation}

    After the PMDC correction obtaining $T_{\mathrm{PMMA}}^{\mathrm{PMDC}}$ and $T_{\mathrm{Al}}^{\mathrm{PMDC}}$, averaged coefficients $\bar{c}_{ij}^{(k)}$ across all pixels were used to compute corrected projections:
    \begin{equation}
    P_{E(k)}^{\mathrm{PMDC}} = \sum_{1 \le i+j \le p} \bar{c}_{ij}^{(k)} (T_{\mathrm{PMMA}}^{\mathrm{PMDC}})^i (T_{\mathrm{Al}}^{\mathrm{PMDC}})^j.
    \label{eq:eq26}
    \end{equation}
    When $p=1$, the model yields an equivalent monoenergetic projection, enabling beam hardening correction. To ensure a fair comparison with other methods that do not perform beam hardening correction, $p=3$ was used to represent the non beam hardening corrected case.
\end{itemize}
\textbf{Note:} FF requires only air scans for calibration. STC uses only PMMA slabs, as PMMA closely approximates the attenuation of soft biological tissue and is commonly used for single-material calibration. ATC and STEPC used PMMA and aluminum slabs for the non-contrast scenario and additional iodixanol slabs for the contrast scenario; PMDC, limited by its two-threshold constraint, uses PMMA and aluminum slabs for both scenarios, with PMMA and aluminum serving as proxies for soft and high-density tissues, respectively, which is the most common choice. This limitation is inherent to PMDC and restricts its applicability to iodine-enhanced imaging, rather than a design choice in our experimental comparison.

\subsection{Metrics}
To assess the proposed correction method, two complementary image quality metrics were employed: the \textit{Mean Local Standard Deviation} (MLSD) for projection-domain uniformity assessment, and the \textit{Ring Artifact Deviation} (RAD) for reconstruction-domain artifact quantification. 

\textbf{(1) Mean Local Standard Deviation (MLSD).} 
For each corrected projection image, projection uniformity is commonly evaluated using variance or standard deviation metrics, as reported in Ref.~\cite{persson2012framework}. However, to properly account for variations in slab phantom thickness across different geometric angles, we adopt the MLSD metric, which quantifies local intensity consistency. Specifically, a sliding window of $20\times20$ pixels was applied across the projection image, and the standard deviation $\sigma_n$ within each window $n$ was calculated. The MLSD was then defined as the mean of all local standard deviations:
\begin{equation}
{\rm MLSD} = \frac{1}{N} \sum_{n=1}^{N} \sigma_n,
\end{equation}
where $N$ denotes the total number of valid local windows. A lower MLSD value indicates higher spatial uniformity of the projection, reflecting more effective correction of detector nonuniformity.

\textbf{(2) Ring Artifact Deviation (RAD).}  
To assess ring artifact suppression in reconstructed images, the Ring Artifact Deviation (RAD) metric provides an effective quantitative measure of ring artifact severity in central uniform regions, as described by Rodesch et al.~\cite{rodesch2023comparison}. As shown in Fig.~\ref{fig:fig5}, each reconstructed slice $I_{\text{real}}(s,x,y)$ was first transformed from Cartesian coordinates $(x,y)$ to polar coordinates $(r,\theta)$ centered on the rotation axis, yielding $I_{\text{polar,real}}(s,r,\theta)$. The angular mean intensity was then computed as:
\begin{equation}
\bar{I}_{\text{polar,real}}(s,r) = \frac{1}{N_\theta} \sum_{\theta=1}^{N_\theta} I_{\text{polar,real}}(s,r,\theta),
\end{equation}
where $N_\theta$ is the total number of angular samples. A second-order polynomial was subsequently fitted along the radial direction to approximate the ideal, ring-free intensity distribution $\bar{I}_{\text{polar,ideal}}(s,r)$.  
The RAD metric was then defined as the standard deviation of the residual intensity differences across all valid radii $r$ and slices $s$:
\begin{equation}
{\rm RAD} = {\rm STD}_{r,s}\!\left\{\bar{I}_{\text{polar,real}}(s,r) - \bar{I}_{\text{polar,ideal}}(s,r)\right\}.
\end{equation}

Lower RAD values correspond to more effective suppression of ring artifacts. Since the single-material and multi-material phantoms possess different internal structures, distinct circular regions of interest (ROIs) with radius $r$ were selected to ensure uniform sampling and fair comparison. For single-material cylindrical phantoms, a diameter of 20~mm was used; for insert phantoms, a diameter of 8~mm was chosen. For in-vivo mouse scans, smaller ROIs were defined due to anatomical heterogeneity: 3.5~mm for the head region and 3.7~mm for the kidney region, as indicated by the red circles in Fig.~\ref{fig:fig12}. The same ROI radius was applied across all compared methods within each dataset to maintain consistency and fairness.

\section{Results}
\label{sec:results}

We evaluated the corrected projection uniformity for different material slabs, and ring artifact suppression in cylindrical phantoms and mouse images. First, simulation experiments were conducted to assess the fitting accuracy of the second-order 2D polynomial model for estimating ideal projections (Fig.~\ref{fig:fig6}, and Table~\ref{tab:table1}). Projection uniformity was then quantitatively evaluated for all PMMA+Al slab combinations (Fig.~\ref{fig:fig7}, \ref{fig:fig8}) and PMMA+Al+iodixanol phantom slabs (Fig.~\ref{fig:fig9}), with averaged MLSD results shown in Table~\ref{tab:table2}. Second, we compared ring artifact suppression qualitatively (Fig.~\ref{fig:fig10},~\ref{fig:fig11},~\ref{fig:fig12}) and quantitatively (Tables~\ref{tab:table3}) for phantoms and mouse imaging. In addition, we demonstrated the potential of the proposed method for beam hardening correction (Fig.~\ref{fig:fig13}) and evaluated its impact on material decomposition accuracy (Fig.~\ref{fig:fig14} and Table~\ref{tab:table4}). Finally, we performed a sensitivity analysis (Fig.~\ref{fig:fig15}, and Table~\ref{tab:table5}) to evaluate the impact of polynomial degree and calibration material choice on different material objects imaging.

\subsection{Ideal Projection Fitting Accuracy Evaluation}

\begin{figure}[!t]
    \centering
    \includegraphics[width=\columnwidth]{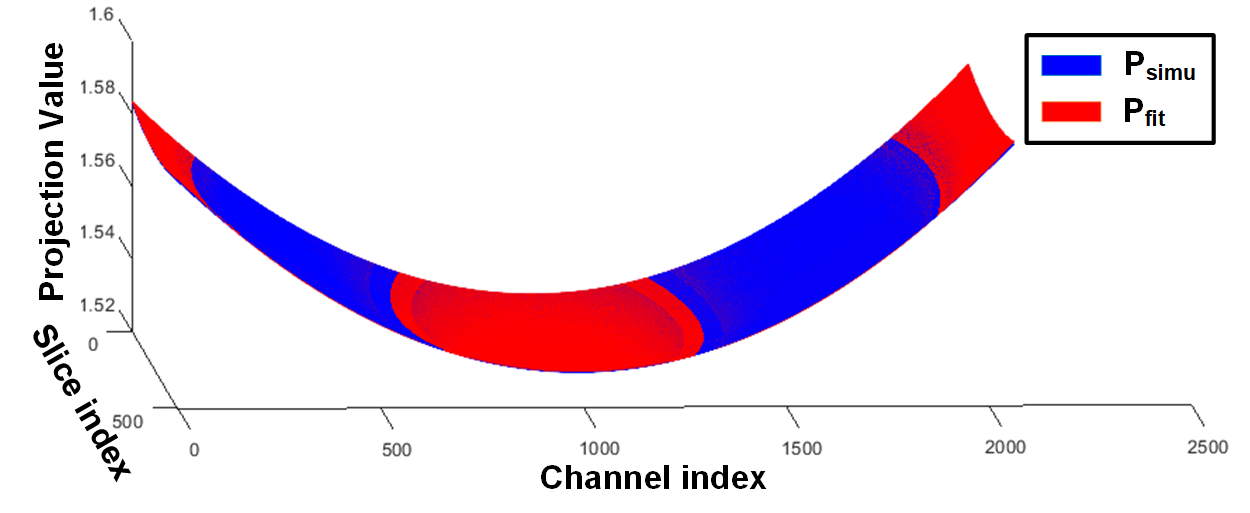}
    \caption{Second-order 2D polynomial fitting results for simulated projection data of a slab phantom (PMMA: 20~mm, Al: 3~mm) under a $0^{\circ}$ tilt angle.}
    \label{fig:fig6}
\end{figure}

\begin{table}[!t]
\centering
\caption{Relative errors of second-order 2D polynomial fitting for simulated projection data.}
\label{tab:table1}
\begin{tabular}{p{0.5cm}p{1.7cm}p{1.4cm}p{1.4cm}p{1.4cm}}
\toprule
\multirow{2}{=}{tilt angles} & \multirow{2}{*}{Combinations} & \multicolumn{3}{c}{RE($\times 10^{-4}$) (MEAN[MIN,MAX])} \\
\cmidrule{3-5}
        &       & Low    & High   & Total \\
\midrule
\multirow{2}{=}{$0^{\circ}$}& PMMA+AL   & 0.82[0.77,0.86]   & 0.81[0.79,0.83]   & 0.80[0.77,0.84] \\
        & PMMA+AL+I  & 0.82[0.79,0.84]  & 0.81[0.79,0.85]  & 0.82[0.78,0.84] \\
\midrule
\multirow{2}{=}{$2.5^{\circ}$}& PMMA+AL   & 0.82[0.76,0.90]   & 0.89[0.87,0.91]   & 0.83[0.77,0.90] \\
        & PMMA+AL+I  & 0.85[0.77,0.89]  & 0.89[0.85,0.91]  & 0.85[0.78,0.89] \\
\midrule
\multirow{2}{=}{$5^{\circ}$}& PMMA+AL   & 1.47[1.08,2.00]   & 1.96[1.78,2.09]   & 1.64[1.32,2.02] \\
        & PMMA+AL+I  & 1.65[1.18,1.92]  & 1.92[1.54,2.08]  & 1.73[1.36,1.96] \\
\bottomrule
\end{tabular}
\end{table}

The proposed STEPC framework employs a second-order 2D polynomial surface to represent the ideal projection. To evaluate the fitting accuracy of this model under different material and thickness combinations of slab phantoms, as well as under varying tilt angles, a series of simulation experiments were conducted.

The X-ray source spectrum was generated using the \textit{Spektr} software~\cite{punnoose2016spektr} with a tube voltage of 80~kV and a 0.5~mm Al filter. The energy response function of the photon-counting detector was adopted from Ref.~\cite{schlomka2008experimental}, with two energy thresholds set at 15~keV and 30~keV. The mass attenuation coefficients of PMMA, Al, and I were obtained from the NIST database~\cite{nist_xray_1995}. The SID and SDD were set to 140~mm and 325~mm, respectively. All thickness and material combinations were identical to those described in Section~\ref{sec:CalibrationSetting}. Three different phantom tilt angles were investigated ($0^{\circ}$, $2.5^{\circ}$, and $5^{\circ}$). Since the primary objective of this experiment was to evaluate the projection fitting accuracy of the second-order 2D polynomial surface, noise effects were not considered. The projection fitting accuracy was quantitatively assessed by computing the relative error (RE) between the fitted ($P_{\mathrm{fit}}$) and simulated projection ($P_{\mathrm{simu}}$) values for different material and thickness combinations:

\begin{equation}
\mathrm{RE}_{s} = \frac{1}{N_{xy}} \sum_{x,y} \frac{\left| P_{\mathrm{fit}}(x,y,s) - P_{\mathrm{simu}}(x,y,s) \right|} {P_{\mathrm{simu}}(x,y,s)},
\end{equation}

Here, $(x,y)$ are the pixel coordinates, $N_{xy}$ is the total number of pixels, and $s$ indexes the calibration combinations of multiple material slabs. As an illustrative example, the fitted and simulated 2D projection images for a slab phantom combination with a PMMA thickness of 20~mm and an Al thickness of 3~mm at a tilt angle of $0^{\circ}$ are shown in Fig.~\ref{fig:fig6}, demonstrating excellent agreement between the two projections. The quantitative results are summarized in Table~\ref{tab:table1}. When no phantom tilt was introduced, the average RE across all energy projections was approximately $0.81\times 10^{-4}$. This value increased only marginally to $0.85\times 10^{-4}$ at a tilt angle of $2.5^{\circ}$, and remained below $2.09\times 10^{-4}$ even at $5^{\circ}$. These results demonstrate that the proposed second-order 2D polynomial surface can accurately estimate ideal projections for slab phantoms with different materials and thicknesses, even in the presence of moderate geometric tilts. Importantly, this approach does not require precise phantom positioning or detailed geometric measurements, thereby significantly simplifying the calibration process.

\subsection{Projection Uniformity for Calibration Slabs}

\begin{figure}[!t]
    \centering
    \includegraphics[width=\columnwidth]{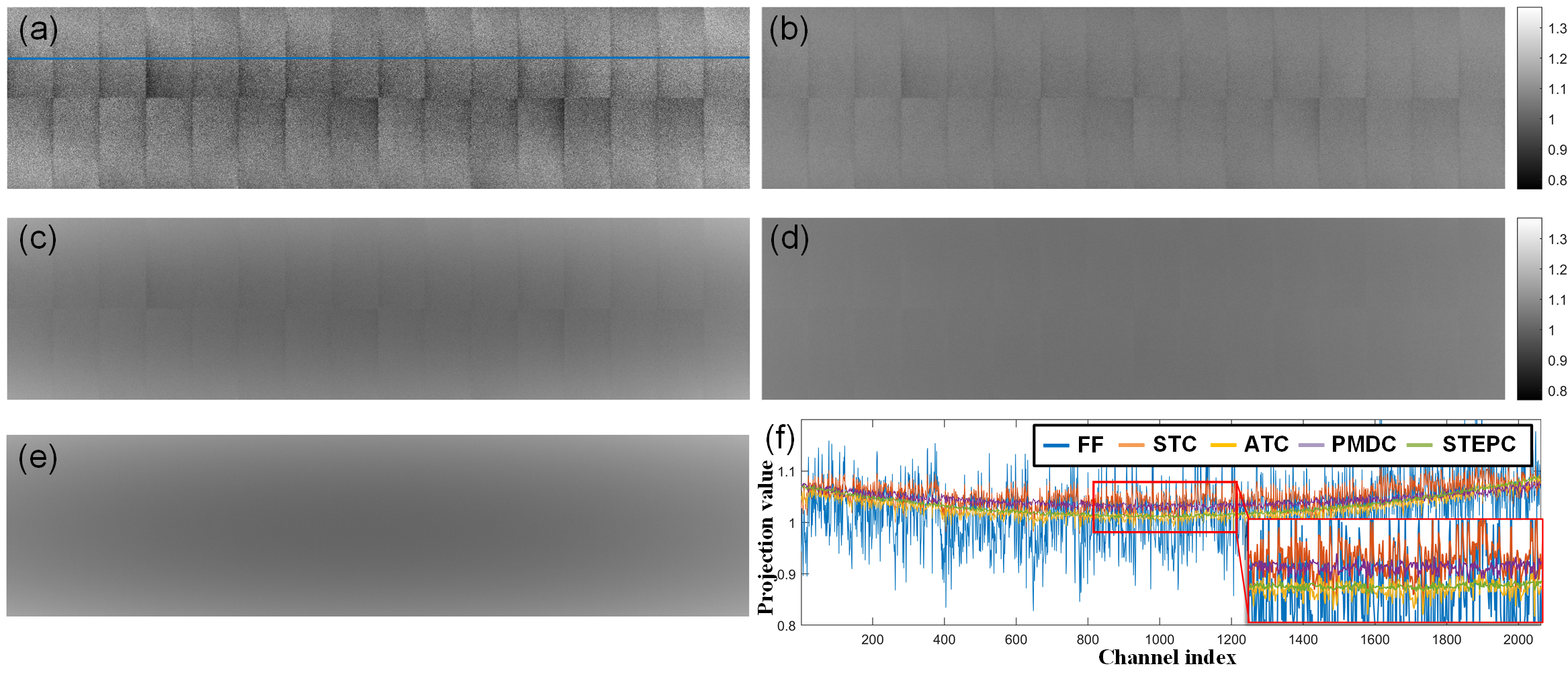}
    \caption{Corrected negative logarithmic projections for the 10 mm PMMA + 1.5 mm Al case using different methods: (a) FF, (b) STC, (c) ATC, (d) PMDC, (e) STEPC, and (f) the corresponding projection profiles, with the blue slice position indicated in (a).}
    \label{fig:fig7}
\end{figure}

We evaluated projection uniformity after applying different correction methods across various combinations of materials and thicknesses. Fig.~\ref{fig:fig7} presents the results for one representative case (10 mm PMMA + 1.5 mm Al). Among all methods, FF correction resulted in the poorest uniformity, while the proposed STEPC method achieved the most consistent and uniform projections.

\begin{figure}[!t]
    \centering
    \includegraphics[width=\columnwidth]{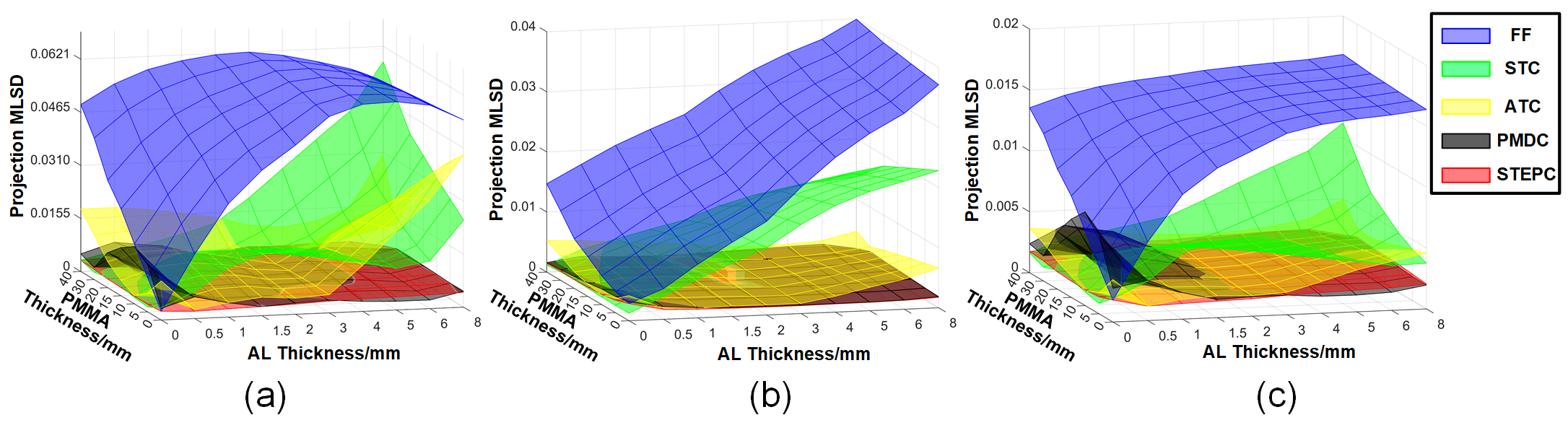}
    \caption{Non-contrast scenario calibration: mean local standard deviation maps of corrected projections for PMMA and aluminum slab combinations using different correction methods: (a) Low, (b) High, (c) Total.}
    \label{fig:fig8}
\end{figure}

\begin{figure}[!t]
    \centering
    \includegraphics[width=\columnwidth]{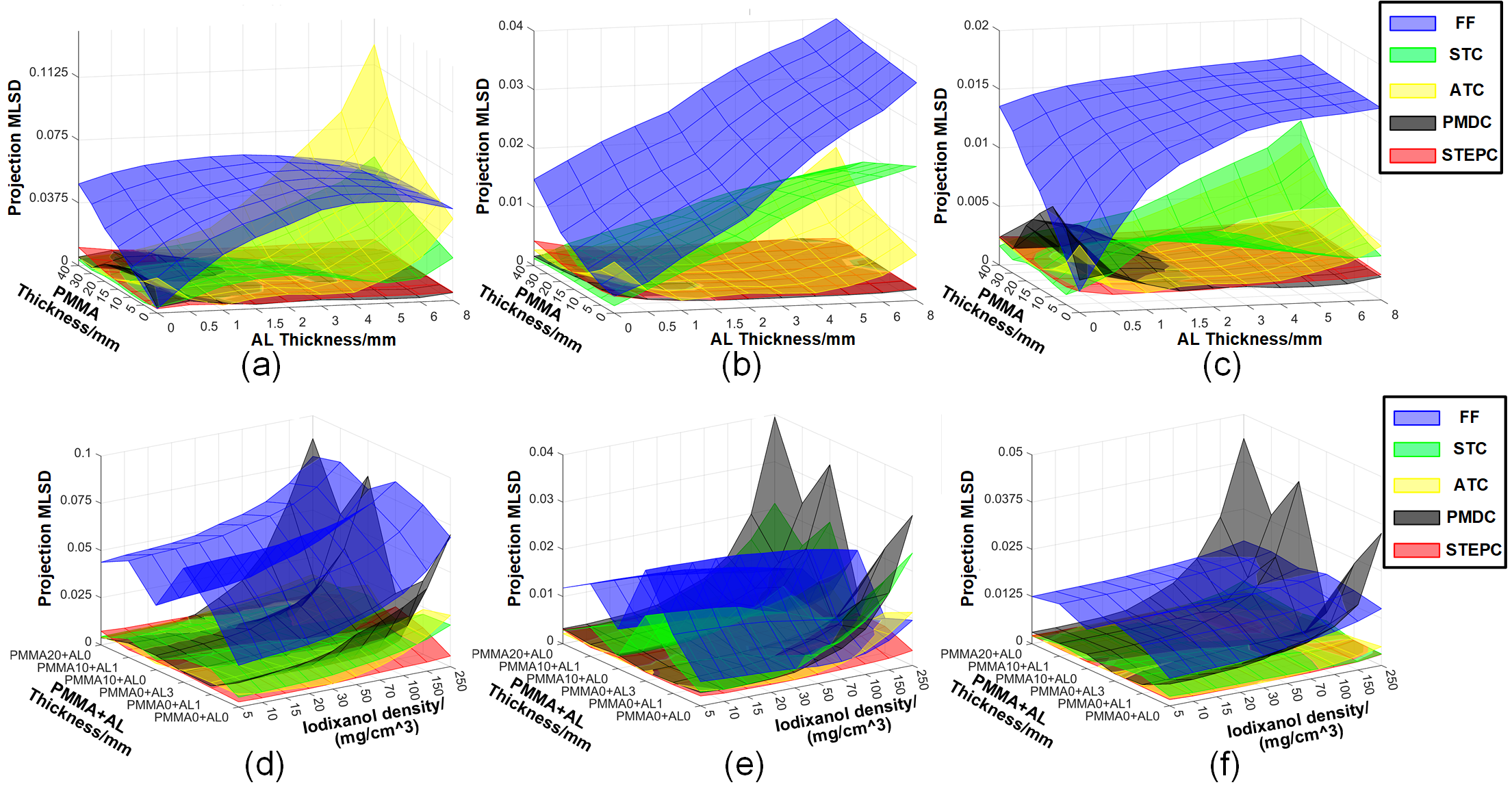}
    \caption{Iodine-enhanced scenario calibration: mean local standard deviation maps of corrected projections for PMMA, aluminum and iodixanol slab combinations using different correction methods: (a) Low of PMMA+AL, (b) High of PMMA+AL, (c) Total of PMMA+AL, (d) Low of PMMA+AL+Iodixanol, (e) High of PMMA+AL+Iodixanol, (f) Total of PMMA+AL+Iodixanol.}
    \label{fig:fig9}
\end{figure}

For the non-contrast scenario calibration, the projection MLSD maps for all PMMA and Al combinations are shown in Fig.~\ref{fig:fig8}. FF consistently performed the worst across all cases. STC performed well with PMMA alone but quickly deteriorated when aluminum was added due to the lack of aluminum calibration. ATC performed well for thin materials but degraded with increasing thickness, especially in the Low and Total energy bins, due to stronger nonlinear effects at lower photon energies. PMDC performed poorly with PMMA-only phantoms, likely due to instability in the material decomposition process. In contrast, STEPC consistently achieved near-best uniformity under all conditions.

For the iodine-enhanced scenario calibration, projection MLSD maps for all PMMA, aluminum, and iodixanol combinations are shown in Fig.~\ref{fig:fig9}. For the only PMMA and Al combinations as shown in Fig.~\ref{fig:fig9}(a-c), the results of FF, STC, and PMDC remain unchanged, as they did not incorporate iodixanol in calibration. Only ATC and STEPC added iodixanol to calibration, resulting in updated results. ATC showed increased MLSD for PMMA+Al projections due to the added material, while STEPC maintained stable performance, with only a slight increase in the PMMA-only projections. For combinations including iodixanol as shown in Fig.~\ref{fig:fig9}(d-f), FF still showed the poorest performance. STC also failed to correct projections containing both aluminum and iodixanol. PMDC's performance degraded sharply with higher iodixanol concentrations. In contrast, ATC performed better overall with the inclusion of all three materials. STEPC again achieved the most uniform projections across all material combinations. Table~\ref{tab:table2} summarizes the average MLSD values. Compared with other methods, STEPC achieves an average MLSD reduction of at least 21.58\%, showing that STEPC consistently yielded the lowest MLSD across both iodine-free and iodine-enhanced scenarios in all energy bins.

\begin{table}[!t]
\centering
\caption{Mean local standard deviation (MLSD) of different methods}
\label{tab:table2}
\begin{tabular}{p{1.7cm}p{0.7cm}cccccc}
\toprule
\multirow{2}{=}{Calibration Materials} & \multirow{2}{*}{Energy} & \multicolumn{5}{c}{MLSD($\times 10^{-2}$)$\downarrow$} \\
\cmidrule{3-7}
        &       & FF    & STC   & ATC   & PMDC  & STEPC \\
\midrule
\multirow{3}{=}{PMMA + AL}& Low   & 5.09   & 1.58   & 1.31  & 0.50  & \textbf{0.38}  \\
        & High  & 2.28  & 1.10  & 0.31  & 0.19  & \textbf{0.18}  \\
        & Total & 1.40  & 0.41  & 0.26  & 0.24  & \textbf{0.19}   \\
\midrule
\multirow{3}{=}{PMMA + AL + Iodixanol} & Low  & 5.31  & 1.09  & 0.90  & 1.81 & \textbf{0.56}  \\
        & High  & 1.13   & 0.76  & 0.35 & 0.75  & \textbf{0.25}  \\
        & Total & 1.34   & 0.29  & 0.24 & 0.82  & \textbf{0.21}  \\   
\bottomrule
\end{tabular}
\end{table}

\subsection{Cylindrical Phantom Imaging}

\begin{figure}[!t]
    \centering
    \includegraphics[width=\columnwidth]{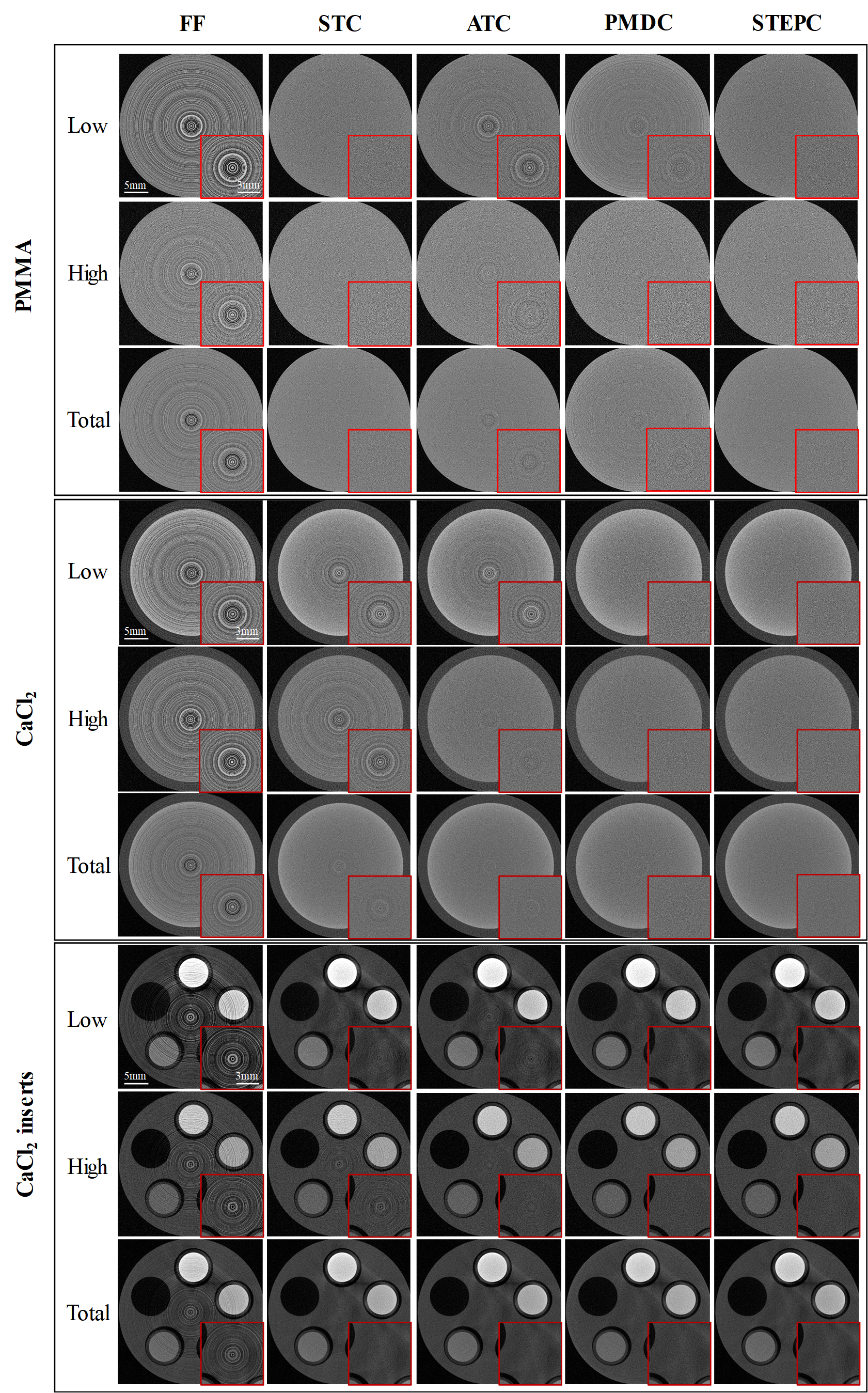}
    \caption{Reconstructed images of the phantom without contrast using different correction methods. Display window: PMMA phantom [-1000, 1000] HU; other phantoms [-1000, 3000] HU. The display window was chosen to account for the high noise level in the images and to better visualize the overall intensity of ring artifacts.}
    \label{fig:fig10}
\end{figure}

Fig.~\ref{fig:fig10} shows reconstructed images of cylindrical phantoms without iodine contrast. For the PMMA cylinder, FF exhibited severe rings, while ATC and PMDC showed mild artifacts, and STC and STEPC produced the least ring artifacts. For the CaCl$_2$ and CaCl$_2$ insert phantom, STEPC and PMDC delivered the best performance, whereas STC and ATC still showed visible artifacts, especially in low-energy images. These observations align with the earlier slab calibration results: STC was calibrated using only PMMA, so it performed well for PMMA but poorly for other materials. PMDC, on the other hand, showed residual artifacts in the PMMA phantom due to decomposition instability when applied to single-material objects. In contrast, STEPC maintained robust performance across both single- and multi-material cases.

\begin{figure}[!t]
    \centering
    \includegraphics[width=\columnwidth]{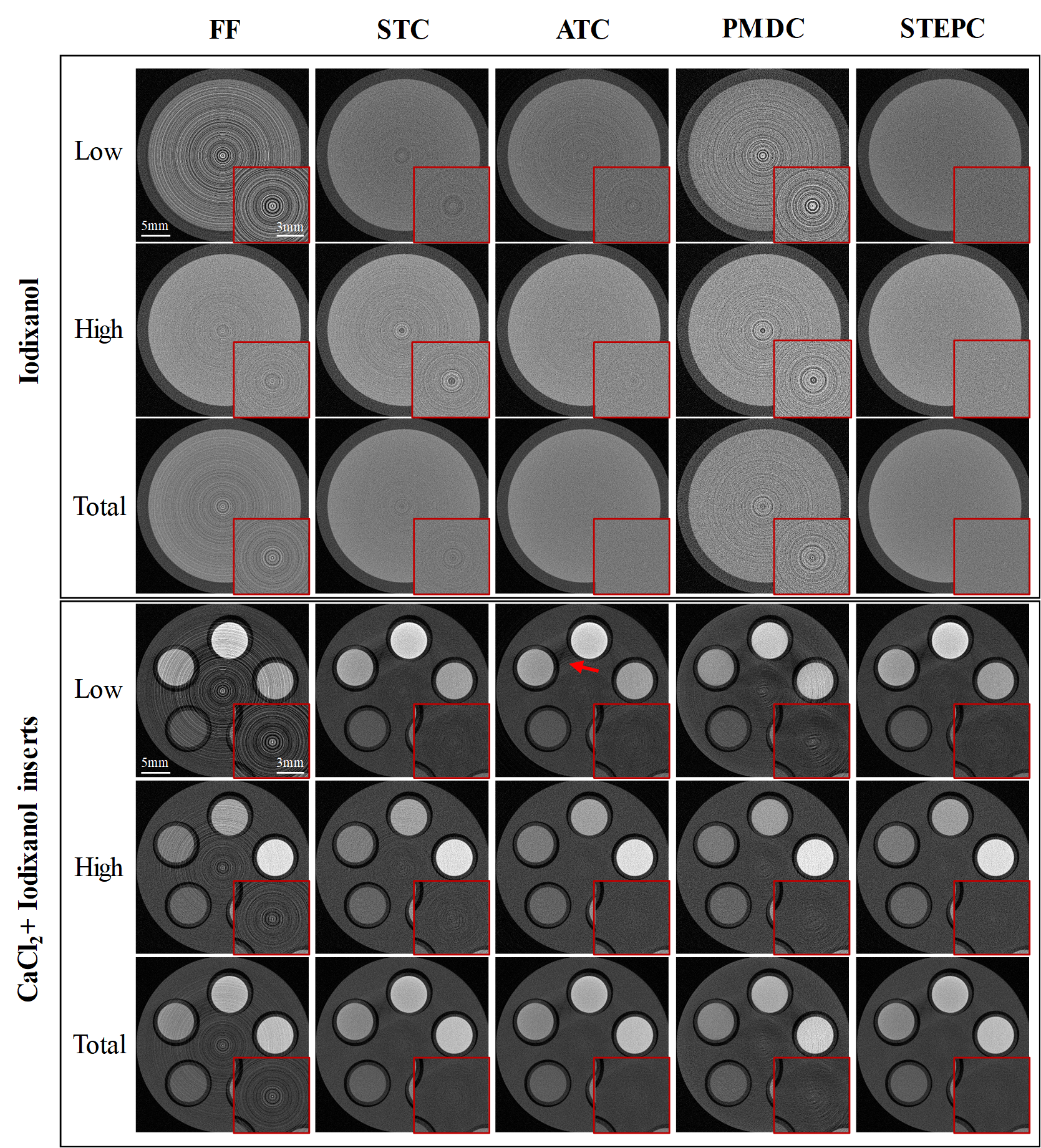}
    \caption{Reconstructed images of the phantom with iodine contrast using different correction methods. Display window: [-1000, 3000] HU.}
    \label{fig:fig11}
\end{figure}

Fig.~\ref{fig:fig11} shows reconstructed images of cylindrical phantoms containing iodine contrast. For the iodixanol-only phantom, FF exhibits severe ring artifacts, particularly in the low-energy bin. STC still shows noticeable artifacts, while ATC leaves only mild residuals. PMDC, however, suffers from prominent ring artifacts. In contrast, STEPC effectively suppresses ring artifacts. For the multi-material inserts (CaCl$_2$ + iodixanol), similar trends are observed. PMDC continues to produce strong artifacts due to the influence of iodine inserts, highlighting its limitation in contrast-enhanced scenarios. Although ATC performs better at the center, visible artifacts remain between the inserts, as indicated by the red arrows. STEPC consistently produces artifact-free images across the entire field, demonstrating its robustness for multi-material correction.

Table~\ref{tab:table3} quantifies the RAD for all phantoms, consistent with the previous qualitative observations. STEPC achieves the lowest RAD in nearly all cases, except for the high-energy PMMA images, where STC performs best. This is likely because STC was calibrated using only PMMA, and PMMA exhibits better linear attenuation at high energie bin, which better matches the linear attenuation assumptions inherent in the STC model.

\begin{table*}[!t]
\centering
\caption{Ring artifact deviation (RAD) of different methods on phantom and mouse images}
\label{tab:table3}
\begin{tabular}{p{1cm}p{1.2cm}p{0.7cm}p{2cm}p{2cm}p{2cm}p{2cm}p{2cm}}
\toprule
\multirow{2}{*}{}                  & \multirow{2}{=}{Objects}      & \multirow{2}{*}{Energy} & \multicolumn{5}{c}{RAD (HU) ↓}            \\ \cmidrule(l){4-8} 
                                        &                                           &                         & FF     & STC            & ATC    & PMDC   & STEPC          \\ \midrule
\multirow{9}{=}{No Contrast Phantoms} & \multirow{3}{=}{PMMA}                     & Low                     & 279.19±81.73$^{**}$ & 27.07±8.63$^{**}$          & 87.87±25.72$^{**}$  & 47.09±12.09$^{**}$  & \textbf{25.99±7.78} \\
                                        &                                           & High                & 135.20±40.65$^{**}$ & \textbf{35.48±10.36}$^{*}$ & 52.13±16.14$^{**}$  & 45.29±13.23$^{**}$  & 35.75±10.23          \\
                                        &                                           & Total                   & 122.26±32.06$^{**}$ & 17.12±4.48$^{*}$          & 29.25±7.20$^{**}$  & 39.17±10.71$^{**}$  & \textbf{16.93±4.5} \\ \cmidrule(l){2-8} 
                                        & \multirow{3}{=}{200mg/ml CaCl$_2$}           & Low                     & 389.03±102.49$^{**}$ & 196.94±67.99$^{**}$         & 168.67±49.17$^{**}$ & 50.43±14.53$^{**}$  & \textbf{39.39±10.72} \\
                                        &                                           & High                    & 419.77±123.76$^{**}$ & 212.37±58.87$^{**}$         & 50.07±13.12$^{**}$  & 34.75±9.60$^{**}$  & \textbf{31.86±8.58} \\
                                        &                                           & Total                   & 170.27±45.48$^{**}$ & 46.70±13.41$^{**}$          & 30.50±8.55$^{**}$  & 37.16±10.28$^{**}$  & \textbf{21.75±5.75} \\ \cmidrule(l){2-8} 
                                        & \multirow{3}{=}{CaCl$_2$ inserts}            & Low                     & 478.42±161.74$^{**}$ & 61.65±16.27$^{**}$          & 80.27±24.74$^{**}$  & 47.46±12.95  & \textbf{47.08±10.63} \\
                                        &                                           & High                    & 285.94±108.10$^{**}$ & 106.85±41.28$^{**}$         & 55.82±18.52$^{**}$  & 46.58±15.42$^{**}$  & \textbf{44.49±14.74} \\
                                        &                                           & Total                   & 196.21±59.77$^{**}$ & 36.02±8.55$^{**}$          & 34.74±8.82$^{**}$  & 33.08±9.21$^{**}$  & \textbf{31.12±7.63} \\ \midrule
\multirow{6}{=}{Contrast Phantoms}    & \multirow{3}{=}{50mg/ml Iodixanol}        & Low                     & 500.56±149.51$^{**}$ & 69.76±20.59$^{**}$          & 66.96±22.93$^{**}$  & 375.98±106.87$^{**}$ & \textbf{34.85±10.62} \\
                                        &                                           & High                    & 116.66±34.29$^{**}$ & 174.31±53.35$^{**}$         & 54.08±15.54$^{**}$  & 272.72±82.96$^{**}$ & \textbf{39.80±11.15} \\
                                        &                                           & Total                   & 167.53±44.08$^{**}$ & 56.71±15.31$^{**}$          & 23.78±6.84$^{**}$  & 218.23±67.21$^{**}$ & \textbf{21.99±6.67} \\ \cmidrule(l){2-8} 
                                        & \multirow{3}{=}{CaCl$_2$ + Iodixanol inserts} & Low                     & 474.48±163.65$^{**}$ & 51.96±15.68$^{**}$          & 39.72±10.45$^{**}$  & 87.50±30.42$^{**}$  & \textbf{34.45±9.12} \\
                                        &                                           & High                    & 221.68±78.25$^{**}$ & 55.61±18.50$^{**}$          & 43.07±13.64$^{**}$  & 65.00±24.72$^{**}$  & \textbf{35.66±12.34} \\
                                        &                                           & Total                   & 193.17±61.15$^{**}$ & 30.19±7.67$^{**}$          & 27.03±6.56$^{**}$  & 56.46±19.37$^{**}$  & \textbf{24.49±6.28} \\ \midrule
\multirow{6}{=}{Mouse}    & \multirow{3}{=}{Head (without contrast)}                & Low                     & 775.55±237.33$^{**}$ & 194.26±55.82$^{**}$         & 164.72±51.56$^{**}$ & 76.67±24.62$^{**}$ & \textbf{62.94±18.39} \\
                                        &                                           & High                    & 373.29±120.95$^{**}$ & 167.71±52.10$^{**}$         & 103.98±32.05$^{**}$ & 89.36±25.86$^{**}$ & \textbf{83.49±24.22} \\
                                        &                                           & Total                   & 339.10±96.45$^{**}$ & 94.25±26.23$^{**}$          & 56.88±17.90$^{**}$  & 56.45±16.39$^{**}$ & \textbf{40.98±12.22} \\ \cmidrule(l){2-8} 
                                        & \multirow{3}{=}{Kidney (iodixanol contrast)} & Low                  & 1116.58±347.04$^{**}$ & 140.50±44.27$^{**}$         & 131.03±40.87$^{**}$ & 283.79±92.77$^{**}$ & \textbf{86.19±26.86} \\
                                        &                                           & High                    & 425.00±127.11$^{**}$ & 177.35±56.45$^{**}$         & 115.76±35.16$^{**}$ & 234.82±84.36$^{**}$ & \textbf{102.59±24.92}\\
                                        &                                           & Total                   & 429.04±128.54$^{**}$ & 83.26±22.70$^{**}$          & 56.13±16.60  & 162.80±47.64$^{**}$ & \textbf{54.91±14.70} \\ \bottomrule
\end{tabular}
\vspace{2mm}
\par\noindent
\footnotesize
\textit{Note:} Statistical significance was evaluated using paired $t$-tests between each baseline method and the proposed method. 
$^{*}: p<0.05$, $^{**}: p<0.001$. Values without superscripts indicate no statistically significant difference.
\end{table*}

\subsection{Mouse Imaging}

\begin{figure}[!t]
    \centering
    \includegraphics[width=\columnwidth]{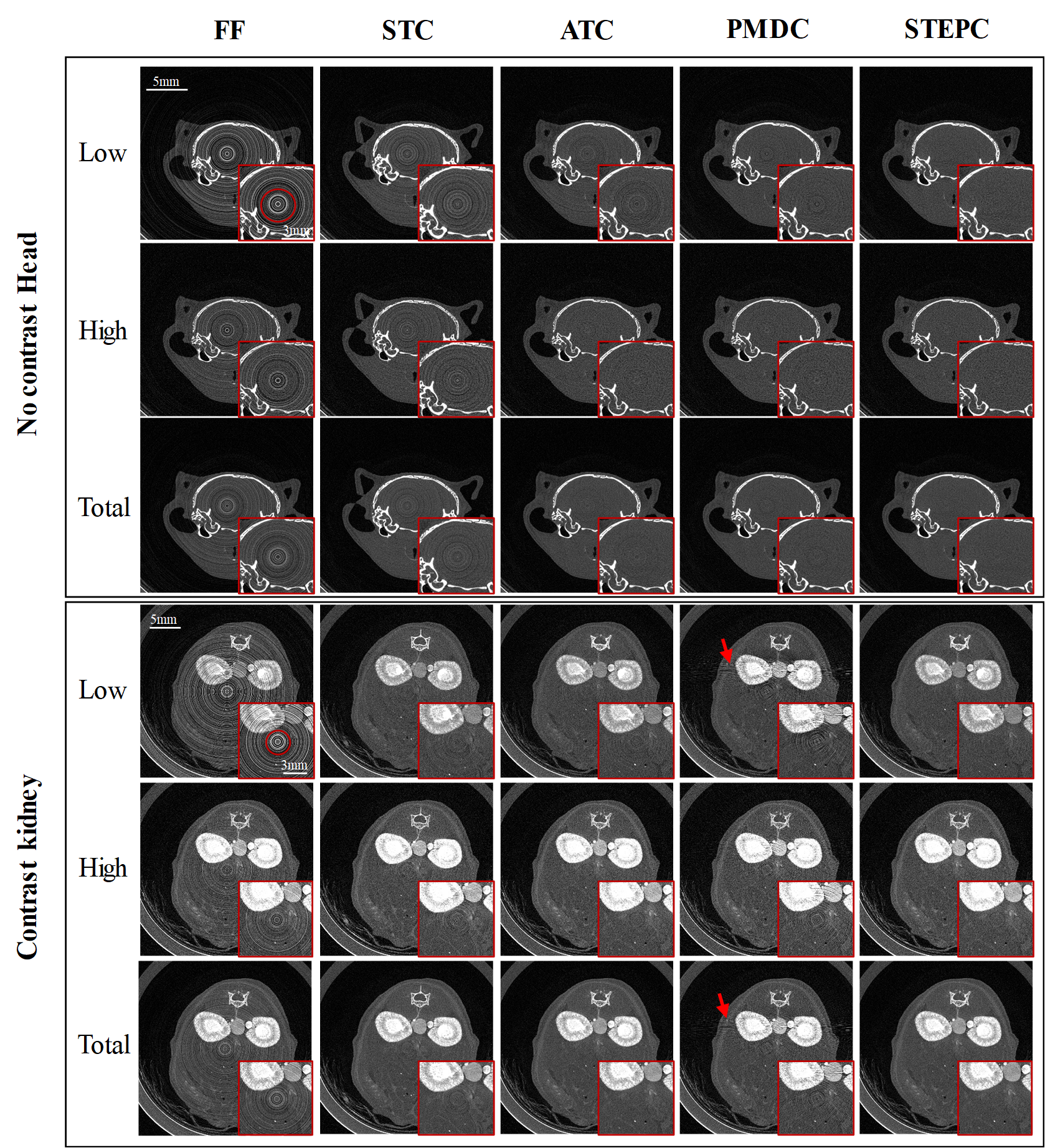}
    \caption{Reconstructed mouse images of the non-contrast head and contrast-enhanced kidney. Display window: [-1000, 3000] HU.}
    \label{fig:fig12}
\end{figure}

Fig.~\ref{fig:fig12} shows reconstructed images of mouse scans, and Table~\ref{tab:table3} includes the corresponding quantitative RAD results. In the non-contrast mouse head, FF produced severe ring artifacts that obscured anatomical details. STC also exhibited visible artifacts, while ATC and PMDC showed relatively milder ring artifacts. However, ATC displayed more pronounced artifacts in the low-energy bin, likely due to its limited ability to correct for the stronger nonlinear effects at lower energies. In contrast, STEPC achieved nearly complete artifact suppression and yielded the lowest RAD values across all energy bins. In the contrast-enhanced kidney images, FF and STC continued to show noticeable artifacts. Both STEPC and ATC performed well due to iodixanol calibration, but ATC showed slight artifacts in the low-energy bin, again likely due to nonlinear spectral effects. PMDC, however, displayed pronounced ring artifacts at the center and severe streak artifacts at the kidney edges, especially in low and total energy images (indicated by red arrows), highlighting its limitation in contrast-enhanced scenarios. STEPC consistently achieved the lowest RAD scores across all energy bins. For both phantom and mouse experiments, STEPC reduced RAD by at least 14.18\% on average. Overall, STEPC consistently delivered the best performance in both non-contrast and contrast-enhanced mouse imaging.

\subsection{Beam hardening correction}

As shown in Fig.~\ref{fig:fig10}, pronounced cupping artifacts were observed in the homogeneous CaCl$_2$ cylindrical phantom, and severe beam hardening artifacts appeared around high-attenuation insert phantom. Based on the proposed STEPC framework, we further extended the method to perform beam hardening correction as described in section \ref{sec:BHcorr}. Three experimental scenarios were tested, including a homogeneous cylindrical phantom, a high-attenuation insert phantom, and a mouse head scan. For comparison, the PMDC method with beam hardening correction (corresponding to $p=1$ in equations~\ref{eq:eq25} and \ref{eq:eq26}) was also included.  

\begin{figure}[!t]
    \centering
    \includegraphics[width=\columnwidth]{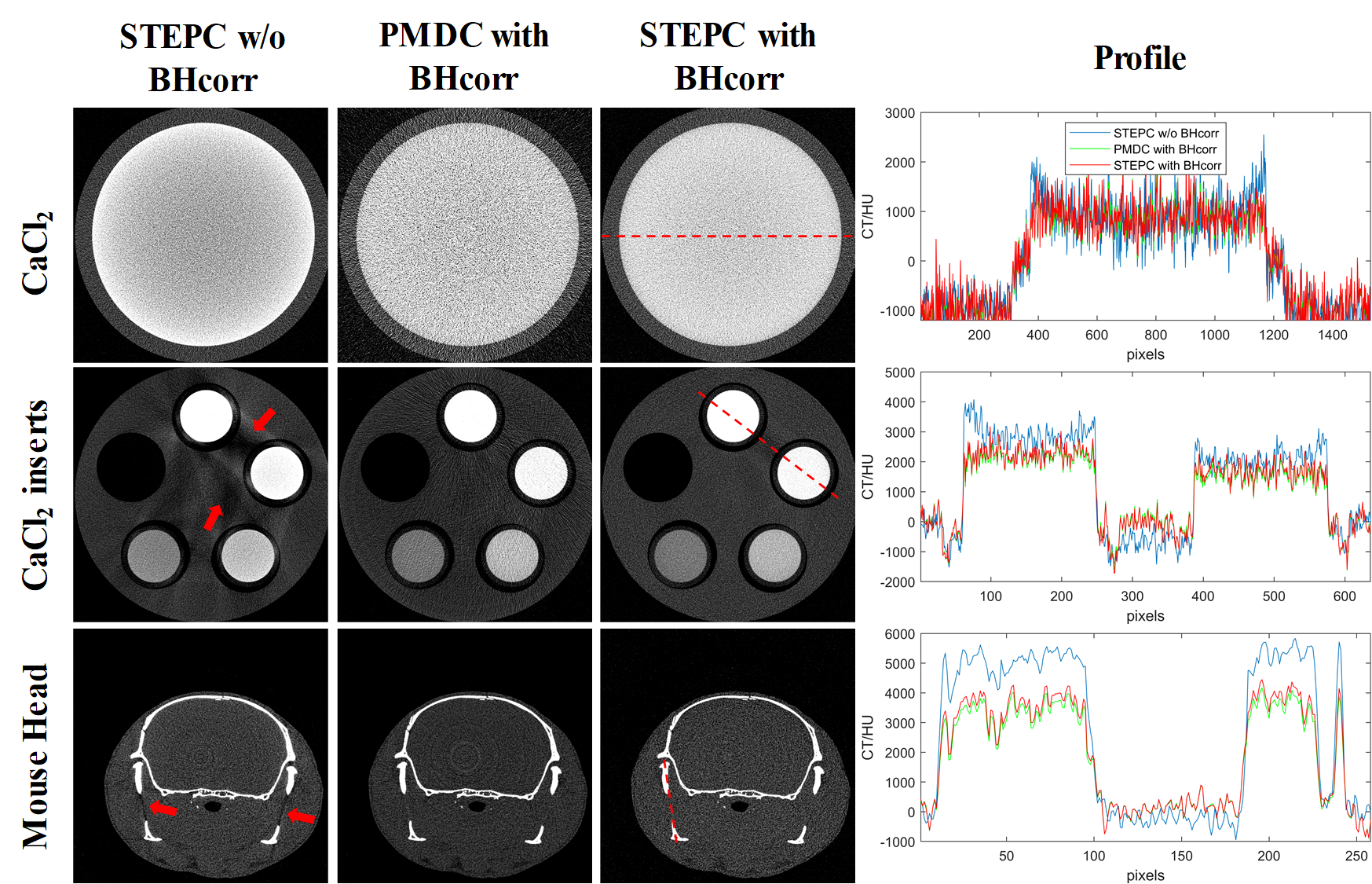}
    \caption{Reconstructed images and corresponding line profiles of the proposed STEPC without and with beam hardening correction, along with PMDC with beam hardening correction, across three experimental scenarios. Display window for the CaCl$_2$ cylindrical phantom: [–500, 1500] HU; others: [-500, 2000].}
    \label{fig:fig13}
\end{figure}

As illustrated in Fig.~\ref{fig:fig13}, both PMDC and STEPC with beam hardening correction effectively suppressed beam hardening artifacts. In the CaCl$_2$ phantom, cupping artifacts were substantially reduced, as evidenced by the flattened intensity profiles. Moreover, the alternating bright and dark streaks between dense inserts and bony structures in the mouse images were significantly mitigated. These results demonstrate that the extended STEPC framework can simultaneously correct detector nonuniformity and beam hardening effects, thereby suppressing both ring and beam hardening artifacts in complex imaging scenarios.

\subsection{Quantitative Evaluation of Material Decomposition}
\label{sec:MDQuantitative}

The quantitative performance of different correction methods in material decomposition was evaluated using CaCl$_2$ and iodixanol insert phantoms. The CaCl$_2$ phantom contained concentrations of 100, 200, 400, and 600~mg/ml, while the iodixanol phantom included inserts of 10, 20, 50, and 100~mg/ml. To reduce the influence of beam hardening artifacts from the CaCl$_2$ phantom on decomposition accuracy, PMDC was evaluated with its intrinsic beam hardening correction (by setting $p=1$ in Eqs.~\ref{eq:eq25} and~\ref{eq:eq26}), whereas FF, STC, ATC, and STEPC were corrected following the procedure in Section~\ref{sec:BHcorr}. After reconstruction, water-CaCl$_2$ and water-iodixanol decompositions were performed using the high and low energy images. The mean and standard deviation of the decomposed concentrations were measured from insert ROIs across all slices. The mean relative error was calculated as:
\begin{equation}
\text{Mean error} = \frac{1}{4} \sum_{i=1}^{4} \frac{\left| C_{\text{real},i} - C_{\text{MD},i} \right|}{C_{\text{real},i}}\times 100\%,
\end{equation}
where $C_{\text{real},i}$ and $C_{\text{MD},i}$ denote the real and decomposed concentrations, respectively.

\begin{figure}[!t]
    \centering
    \includegraphics[width=\columnwidth]{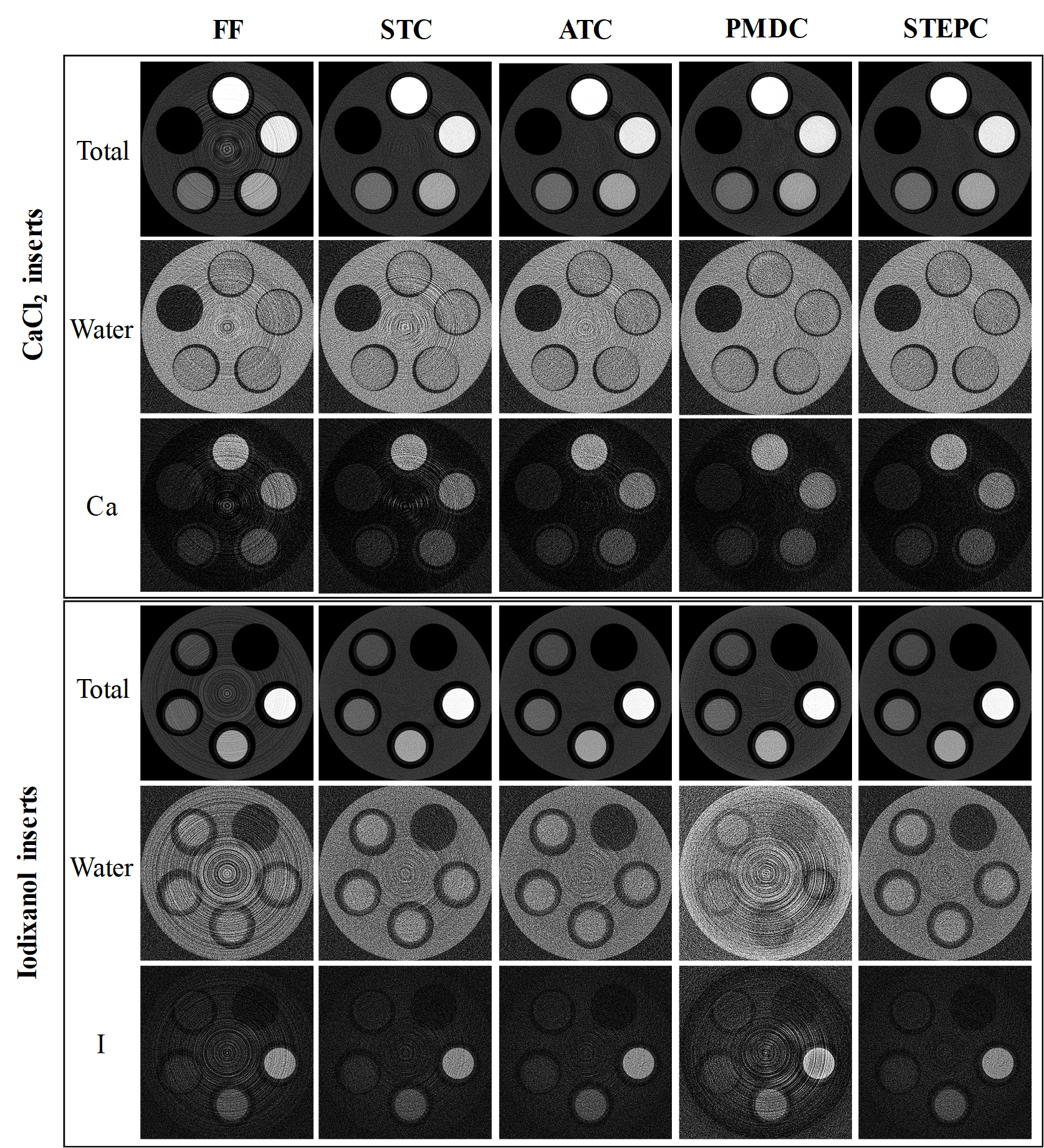}
    \caption{Material decomposition results for the CaCl$_2$ and iodixanol insert phantoms. The top row shows the total-energy bin images, while the middle and bottom rows display the water-CaCl$_2$ and water-iodixanol basis material maps, respectively. Display windows: total-energy bin image: [-500,2000]~HU; water map: [0,2000]~mg/ml; Ca map: [0,1000]~mg/ml; I map: [0,200]~mg/ml.}
    \label{fig:fig14}
\end{figure}

The quantitative results are summarized in Table~\ref{tab:table4}, and the corresponding material decomposition maps are shown in Fig.~\ref{fig:fig14}. For CaCl$_2$ decomposition, all methods except FF achieved mean errors below 1\%. PMDC yielded the smallest mean error and standard deviation, while the proposed STEPC achieved comparable accuracy with only marginally higher values, remaining well within the acceptable quantitative range. For iodixanol decomposition, PMDC exhibited the largest mean error and deviation, particularly at low concentrations, whereas STEPC achieved the smallest mean error and standard deviation among all methods. As shown in Fig.~\ref{fig:fig14}, residual ring artifacts of varying severity were still visible in the decomposed basis images, potentially affecting quantitative accuracy. However, since the mean concentrations were measured within insert ROIs, the impact of mild artifacts on quantitative metrics was minimal. Pronounced artifacts, such as those observed in FF for CaCl$_2$ and in PMDC for iodixanol, led to increased errors and standard deviations. Overall, these results demonstrate that the proposed STEPC framework effectively suppresses artifacts while maintaining, or even slightly improving the quantitative accuracy of material decomposition.

\begin{table*}[!t]
\centering
\caption{Quantitative accuracy analysis of material decomposition for CaCl$_2$ and iodixanol phantoms across~different~correction~methods.}
\label{tab:table4}
\begin{tabular}{p{1cm}p{2.5cm}p{2cm}p{2cm}p{2cm}p{2cm}p{2cm}}
\toprule
\multirow{2}{*}{Phantom}           & \multirow{2}{*}{Concentration (mg/ml)} & \multicolumn{5}{c}{Material decomposition concentration (mg/ml)}                                              \\ \cmidrule(l){3-7} 
                                   &                                        & FF                     & STC                    & ATC                  & PMDC                & STEPC                \\ \midrule
\multirow{5}{=}{CaCl$_2$ inserts}     & 100                                    & 94.64±269.30           & 98.12±244.69           & 98.11±257.41         & 100.99±233.10       & 98.83±254.04         \\ 
                                   & 200                                    & 203.16±333.08          & 201.23±298.37          & 200.80±306.54        & 197.44±292.60       & 201.79±308.16        \\
                                   & 400                                    & 404.94±401.08          & 403.90±363.53          & 402.27±362.82        & 402.44±359.95       & 403.91±368.90        \\
                                   & 600                                    & 594.67±430.73          & 599.06±401.91          & 598.43±392.88        & 598.92±389.63       & 596.86±395.41        \\
                                   & mean error/std                         & 2.27\%/358.55          & 0.91\%/327.13          & 0.78\%/329.91        & 0.77\%/318.82       & 0.89\%/331.63      \\ \midrule
\multirow{5}{=}{Iodixanol inserts} & 10                                     & 10.03±42.82            & 9.97±38.00             & 10.11±38.27          & 4.30±65.00          & 10.01±38.55          \\  
                                   & 20                                     & 20.12±44.74            & 20.07±39.28            & 20.11±39.50          & 22.17±63.41         & 20.08±39.95          \\
                                   & 50                                     & 49.78±49.24            & 49.86±42.28            & 49.82±42.29          & 57.26±65.80         & 49.89±42.09          \\
                                   & 100                                    & 100.01±55.51           & 100.02±46.73           & 100.04±46.62         & 96.99±80.14         & 100.04±44.92         \\
                                   & Mean error/std                         & 0.34\%/48.08           & 0.25\%/41.57           & 0.52\%/41.67         & 21.33\%/68.58       & 0.21\%/41.38       \\ \bottomrule
\end{tabular}
\end{table*}

\subsection{Sensitivity Analysis}

\begin{figure}[!t]
    \centering
    \includegraphics[width=7cm]{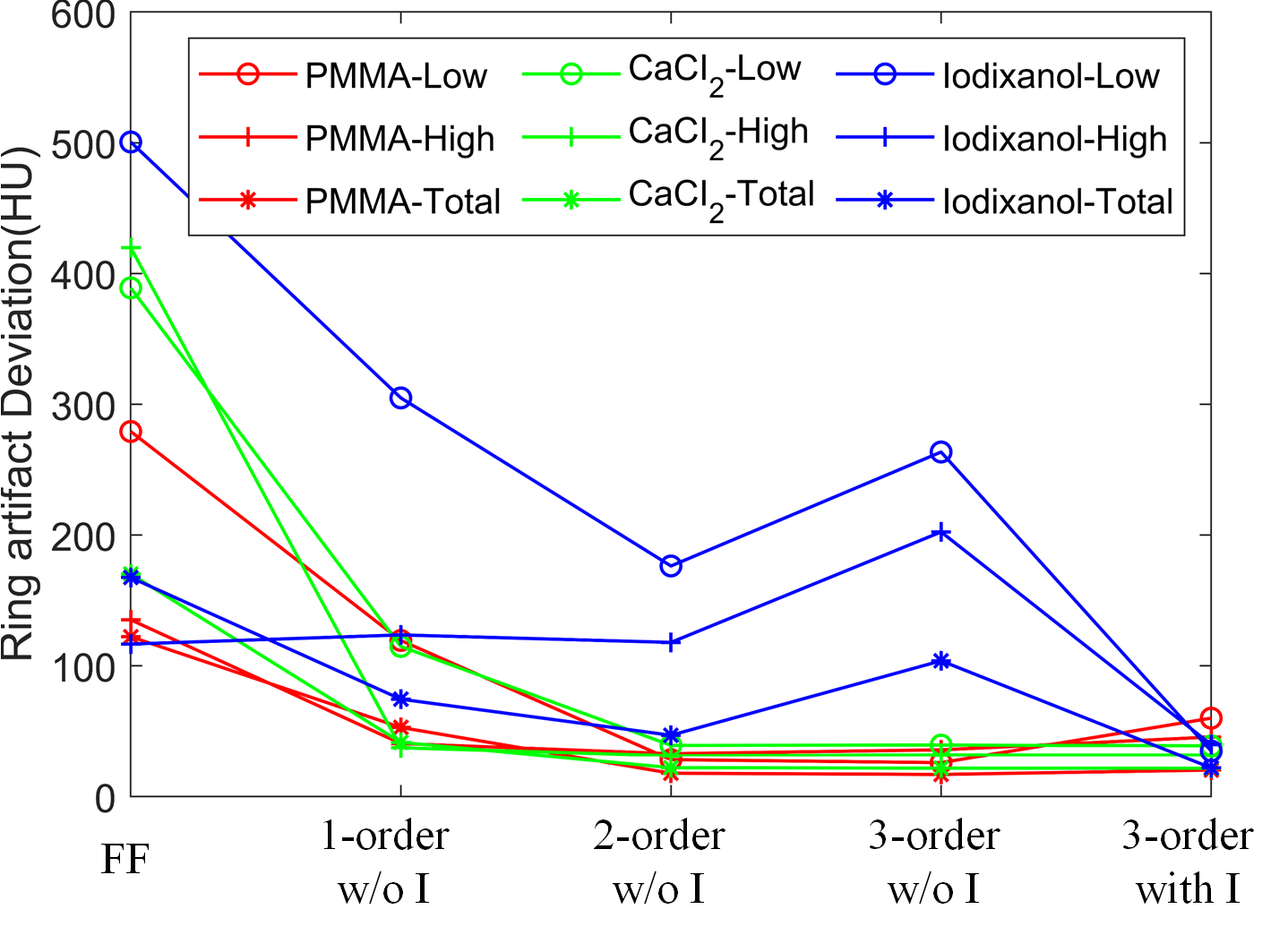}
    \caption{Sensitivity analysis of STEPC with respect to polynomial order and calibration materials.}
    \label{fig:fig15}
\end{figure}

We evaluated the sensitivity of STEPC to polynomial order and calibration material selection, as shown in Fig.~\ref{fig:fig15} and Table~\ref{tab:table5}. When using only PMMA and aluminum for calibration (labeled ``w/o I''), RAD for PMMA and CaCl$_2$ phantoms dropped quickly at orders below 2 but showed little improvement or even slight degradation at order 3, such as for PMMA at high energy. For the iodixanol phantom, RAD decreased up to second order but increased at third order, likely due to overfitting to PMMA and aluminum, reducing generalization to new materials. Therefore, a second-order polynomial is typically sufficient.

\begin{table}[!t]
\centering
\scriptsize  
\caption{Quantitative evaluation of STEPC with respect to polynomial order and calibration materials.}
\label{tab:table5}
\begin{tabular}{p{1.4cm}p{0.35cm}p{0.35cm}p{0.35cm}p{0.35cm}p{0.35cm}p{0.35cm}p{0.35cm}p{0.35cm}p{0.35cm}}
\toprule
\multirow{3}{*}{Setting}   & \multicolumn{9}{c}{Ring Artifact Deviation (HU) ↓}                                                                                       \\ \cmidrule(l){2-10} 
                       & \multicolumn{3}{c}{PMMA}                        & \multicolumn{3}{c}{200mg/ml CaCl$_2$}               & \multicolumn{3}{c}{50mg/ml Iodixanol}            \\
                       & Low            & High           & Total         & Low            & High           & Total          & Low            & High           & Total          \\ \midrule
FF                     & 279.19         & 135.20          & 122.26         & 389.03         & 419.77         & 170.27         & 500.56         & 116.66          & 167.53         \\
1-order w/o I  & 119.40          & 40.55          & 52.88         & 114.80          & 37.32          & 41.47          & 304.87         & 123.63          & 74.50          \\
2-order w/o I  & 28.27          & \textbf{32.81} & 17.86         & 39.19          & \textbf{31.16}   & 22.12          & 176.35          & 117.92          & 46.91          \\
3-order w/o I  & \textbf{25.99} & 35.75          & \textbf{16.93} & 39.39          & 31.86          & 21.75          & 263.50         & 202.38         & 103.85          \\
3-order with I & 59.98          & 45.56          & 20.31         & \textbf{38.87} & 31.63          & \textbf{21.63} & \textbf{34.85} & \textbf{39.80} & \textbf{21.99} \\ \bottomrule
\end{tabular}
\end{table}

When iodixanol slabs were included in calibration (labeled ``with I''), RAD for iodixanol phantoms decreased significantly, indicating improved correction. However, RAD for PMMA slightly increased, and CaCl$_2$ remained stable or improved slightly at low and total energies. This is because adding iodixanol forces the model to learn a more balanced representation across all three materials. Since iodixanol and PMMA have similar attenuation distribution and differ more from aluminum (see Fig.~\ref{fig:fig1}), iodixanol reduces the model's specificity to PMMA. However, this impact is limited in practice. Firstly, real biological imaging typically involves multiple materials (e.g., soft tissue, bone, and contrast agents), where the correction remains stable and effective for multi-component objects. More importantly, we typically know whether the contrast agent is present. For non-contrast cases, a model calibrated only with PMMA and aluminum can be used; for contrast-enhanced scans, adding contrast agents into the calibration process can significantly improve correction performance. In summary, a second-order polynomial is sufficient for STEPC, and including contrast materials in calibration is beneficial for contrast-enhanced imaging.

\section{DISCUSSION}
\label{sec:discussion}

In this study, we proposed the Signal-to-Nonuniformity  Error Polynomial Calibration (STEPC) method to address detector response nonuniformity in photon-counting CT systems. It begins by generating ideal flat-field projections using a 2D second-order polynomial fit. An empirical polynomial model is then built on the residuals across all energy thresholds. This approach overcomes the limitation of requiring more energy thresholds for multi-material calibration, enabling accurate prediction and correction of nonuniformity errors under complex incident spectra.

Detector nonuniformity is dependent on the incident X-ray spectra as indicated in Eq.~\ref{eq:eq12}. Previous studies have used slab phantoms with varying thicknesses of PMMA (to simulate soft tissue) and aluminum (to simulate bone) to perform single or dual-material calibration. However, no calibration strategies have yet been developed specifically for contrast agents. We found that the choice of calibration materials is critical. As shown in Fig.~\ref{fig:fig10}, STC calibrated only with PMMA performs well for PMMA but poorly for CaCl$_2$ phantom. Similarly, PMDC, calibrated with both PMMA and aluminum, achieves better correction for both PMMA, CaCl$_2$ phantoms and mouse head, but its performance degrades in the presence of contrast agents, as seen in Fig.~\ref{fig:fig11} and Fig.~\ref{fig:fig12}. Notably, nonlinear spectral effects are more prominent at lower energy bins due to increased beam hardening and material-specific absorption variations. This significantly challenges calibration models that assume linearity, such as ATC, which exhibits more severe ring artifacts in low-energy images, as shown in Fig.~\ref{fig:fig10} - \ref{fig:fig12}. To address these challenges, we designed a dedicated slab phantom containing an iodinated contrast solution to calibrate contrast-enhanced objects. However, most photon-counting detectors are currently limited to two energy thresholds, making material decomposition methods like PMDC unsuitable for three-material decomposition in the projection domain. The proposed STEPC method overcomes this limitation by directly modeling and predicting nonuniformity errors without relying on material decomposition and additional energy thresholds. By employing a nonlinear multi-energy polynomial model, STEPC effectively captures spectral nonlinearity and enables more accurate and robust correction in complex multi-material imaging scenarios.

On the other hand, the proposed STEPC method can be flexibly adapted to different imaging scenarios. For non-contrast-enhanced mouse scans, using only PMMA and aluminum in calibration is sufficient to approximate the possible incident spectra, as shown in Fig.~\ref{fig:fig1}. For contrast-enhanced scans, only an additional slab phantom containing the contrast agent needs to be included, no changes to the model itself are required. STEPC also demonstrates strong robustness, successfully suppressing ring artifacts across single-material, dual-material, and even triple-material phantoms with iodinated contrast, as shown in Fig.~\ref{fig:fig10} and Fig.~\ref{fig:fig11}. In vivo results also confirm its superior performance in both non-contrast mouse head and contrast-enhanced kidney imaging (Fig.~\ref{fig:fig12}). Moreover, STEPC can be readily extended for beam hardening correction, enabling simultaneous suppression of both ring and beam-hardening artifacts, as demonstrated in Fig.~\ref{fig:fig13}. The beam-hardening correction module can also operate independently or be integrated with other calibration methods such as FF, STC, or ATC (Fig.~\ref{fig:fig14}), making the framework applicable to more complex imaging conditions. Regarding the quantitative impact on material decomposition, we evaluated the accuracy of calcium and iodine quantification. Although STEPC did not achieve the best precision for calcium, the quantitative error remained small and within acceptable limits, while iodine quantification was the most accurate among all methods.

Despite its advantages, STEPC may be susceptible to overfitting due to the use of higher-order polynomial models. Overfitting mainly arises from noise and material variability. To suppress noise-induced overfitting, sufficient frame averaging is required during calibration; in this work, 600 frames were acquired for each slab combination and averaged for model fitting. Material overfitting depends on the selection and combinations of calibration materials. For biological imaging, appropriate combinations of PMMA and aluminum slabs with different thicknesses should be selected to effectively cover typical attenuation ranges, as illustrated in Fig.~\ref{fig:fig1}. When the number of calibration combinations needs to be reduced, overfitting risk can be mitigated by selecting thickness combinations that yield a broad and well-distributed range of attenuation values. However, for materials not included in calibration, such as iodine, whose attenuation characteristics differ substantially from those of PMMA and aluminum, correction performance may degrade, particularly for higher-order models, as shown in Table~\ref{tab:table5} and Fig.~\ref{fig:fig15}. In such cases, incorporating additional calibration materials can improve robustness for more diverse imaging scenarios. Empirically, second-order model already achieves effective correction, with only marginal improvement from higher-order models. Nevertheless, for higher-order polynomials or limited calibration data, overfitting may still occur, and additional strategies such as parameter regularization, cross-validation on a subset of calibration slabs, or averaging over more frames could be considered in future implementations. Consequently, limiting the polynomial order to no higher than third order is recommended to balance correction accuracy, robustness, and computational efficiency.

All experiments in this study were conducted on a single custom Micro-PCCT system. In principle, however, STEPC can be readily extended to other Micro-PCCT systems and even to clinical PCCT platforms. Different systems may exhibit distinct detector designs and spectral characteristics, and additional system-specific corrections may be required prior to applying STEPC. For example, clinical PCCT systems typically operate under higher X-ray flux conditions, where pile-up effects become non-negligible and should be corrected before nonuniformity calibration. Changes in energy threshold settings, tube voltage, or filtration conditions modify the effective X-ray spectrum and detector response, and therefore recalibration is required to maintain correction accuracy, which is standard practice for PCCT calibration methods. Regarding robustness, the ideal reference projections in STEPC are generated via second-order 2D polynomial surface and multi-frame averaging across all detector pixels, which mitigates statistical noise and avoids introducing spatial bias. As demonstrated in Figs.~\ref{fig:fig10}--\ref{fig:fig12}, STEPC does not amplify noise or induce center-to-edge artifacts under the studied acquisition conditions. However, for more complex system configurations, such as those employing bowtie filter or operating in extremely low-count regimes, noise propagation and model robustness may require additional consideration. In such cases, additional regularization strategies or increased calibration averaging may be beneficial to preserve correction stability. For other contrast agents, such as gadolinium, STEPC can in principle be applied following the same calibration procedure used for iodine-based imaging. In addition, when a long-time interval exists between calibration and data acquisition, detector drift and small variations in the X-ray spectrum may degrade correction performance, which is a common issue in photon-counting CT systems~\cite{schmidt2017spectral}. Periodic recalibration, for example every few months, is therefore recommended to maintain correction stability.

Future work may focus on the following directions:
\begin{enumerate}
    \item \textbf{Optimization of Calibration Materials:} The current use of various thickness or density combinations increases calibration complexity. Future work could optimize and reduce the number of combinations to simplify the calibration process.

    \item \textbf{Extension to Complex Scenarios:} This study focuses on dual-threshold Micro-PCCT systems. As next-generation photon-counting detectors support more energy thresholds, STEPC could be generalized to higher-dimensional spectral spaces to better handle multi-contrast-agent imaging. It is also necessary to assess the generalizability of iodine-based calibration to other contrast agents (e.g., gadolinium or barium).

    \item \textbf{Neural Network Models:} STEPC uses empirical polynomial model for errors prediction. While efficient and interpretable, it may underperform in more complex scenarios involving multiple contrast agents or high noise levels. Future work could incorporate neural networks to improve correction accuracy and generalizability.
\end{enumerate}

\section{Conclusion}
\label{sec:conclusion}

In this study, we proposed the STEPC framework to address measurement nonuniformity of PCDs in Micro-PCCT systems under multi-material imaging conditions. STEPC treats detector nonuniformity as a measurable system error and employs a multi-energy polynomial formulation to implicitly incorporate both non-ideal detector response and material-dependent attenuation effects, enabling accurate prediction and correction of nonuniformity errors in the projection domain. By calibrating the model using multi-material slab phantoms, the proposed framework achieves effective correction for complex multi-material objects, even when only two energy thresholds are available. Experimental results demonstrate that, compared with existing single- and multi-energy correction approaches, STEPC more effectively exploits spectral measurement information, leading to improved projection uniformity and substantial suppression of ring artifacts in both multi-material phantom experiments and mouse imaging with and without contrast agents. In addition, the framework can be naturally extended to compensate for beam hardening effects. Overall, STEPC provides a flexible and robust calibration strategy for modeling and correcting PCD measurement nonuniformity, making it a potentially general-purpose calibration framework for Micro-PCCT systems.

\appendices

\section{Derivation of Eq.~\eqref{eq:eq12}}
\label{app:derivation}

The pixel-wise nonuniformity error can be expressed as:
\begin{equation}
N_{k,j}^{\mathrm{error}} = N_{k,j} - N_{k,j}^{\mathrm{ideal}}
\label{eq:eq31}
\end{equation}

According to Eq.~\eqref{eq:eq11}, the ideal photon counts can be rewritten as
\begin{equation} 
    \begin{split} 
        &N_{k,j}^{\mathrm{ideal}} = \\ 
        &\int_{E_{k}-\Delta E_{k,j}}^{E_{\max}} \int_0^{E_{\max}} S(E) e^{-\int_{l}{\mu(E, \vec{x}) dl}} R_j^{\mathrm{ideal}}(E', E) \, dE \, dE' \\ 
        &- \int_{E_{k}-\Delta E_{k,j}}^{E_{k}} \int_0^{E_{\max}} S(E) e^{-\int_{l}{\mu(E, \vec{x}) dl}} R_j^{\mathrm{ideal}}(E', E) \, dE \, dE' 
    \end{split} 
    \label{eq:eq32} 
\end{equation}

Substituting Eq.~\eqref{eq:eq10} and Eq.~\eqref{eq:eq32} into Eq.~\eqref{eq:eq31}, the nonuniformity error can be expressed as
\begin{equation} 
    \begin{split} 
        &N_{k,j}^{\mathrm{error}} = \\ 
        &\int_{E_{k}-\Delta E_{k,j}}^{E_{\max}} \int_0^{E_{\max}} S(E) e^{-\int_{l}{\mu(E, \vec{x}) dl}} \Delta R_j(E', E) \, dE \, dE' \\ 
        &+ \int_{E_{k}-\Delta E_{k,j}}^{E_{k}} \int_0^{E_{\max}} S(E) e^{-\int_{l}{\mu(E, \vec{x}) dl}} R_j^{\mathrm{ideal}}(E', E) \, dE \, dE' 
    \end{split} 
\end{equation}

In practical photon-counting detectors, the threshold offset $\Delta E_{k,j}$ is typically small compared with the energy bin width. Under this assumption, the above expression can be approximated as:
\begin{equation} 
    \begin{split} 
        &N_{k,j}^{\mathrm{error}} \approx \\ 
        & \int_{E_{k}}^{E_{\max}} \int_0^{E_{\max}} S(E) e^{-\int_{l}{\mu(E, \vec{x}) dl}} \Delta R_j(E', E) \, dE \, dE' \\ 
        &+ \Delta E_{k,j} \int_{0}^{E_{\max}} S(E)\, e^{-\int_{l}\mu(E,\vec{x})\,dl}\, R_j^{\mathrm{ideal}}(E_{k},E)\, dE. 
    \end{split} 
    \label{eq:eq12_app}
\end{equation}

Eq~\eqref{eq:eq12_app} corresponds to the approximate nonuniformity error model given in Eq.~\eqref{eq:eq12} of the main text.

\bibliographystyle{IEEEtran}
\bibliography{ref}

\end{document}